\documentclass[letterpaper]{JHEP3}
\usepackage{amssymb,amsfonts}
\usepackage{cite}
\usepackage{epsfig}

\newcommand{\bra}[1]{\langle #1|}
\newcommand{\ket}[1]{|#1\rangle}

\newcommand{\Z}{{\bf Z}}
\newcommand{\BU}{{\bf U}}
\newcommand{\CD}{{\cal D}}
\newcommand{\CF}{{\cal F}}

\newcommand{\CI}{{\cal I}}
\newcommand{\CJ}{{\cal J}}

\newcommand{\CT}{{\cal T}}
\newcommand{\CV}{{\cal V}}

\newcommand{\CDX}{{\cal D}_{\!\!_{X^+}}}

\newcommand{\p}{\partial}

\renewcommand{\bar}[1]{\overline{#1}}
\renewcommand{\tilde}[1]{\widetilde{#1}}
\newcommand{\be}{\begin{equation}}
\newcommand{\ee}{\end{equation}}
\newcommand{\bea}{\begin{eqnarray}}
\newcommand{\eea}{\end{eqnarray}}

\usepackage{graphicx}
\usepackage{latexsym}
\title{M-Theory Through the Looking Glass:\\
Tachyon Condensation in the $E_8$ Heterotic String}
\author{Petr Ho\v{r}ava and Cynthia A. Keeler\\
Berkeley Center for Theoretical Physics and Department of Physics\\
University of California, Berkeley, CA, 94720-7300\\
and\\
Theoretical Physics Group, Lawrence Berkeley National Laboratory\\
Berkeley, CA 94720-8162, USA}
\abstract{We study the spacetime decay to nothing in string theory and 
M-theory.  First we recall a nonsupersymmetric version of heterotic M-theory,
in which bubbles of nothing -- connecting the two $E_8$ boundaries by a throat
-- are expected to be nucleated.  We argue that the fate of this system should
be addressed at weak string coupling, where the nonperturbative instanton
instability is expected to turn into a perturbative tachyonic one.
We identify the unique string theory that could describe this process:  The
heterotic model with one $E_8$ gauge group and a singlet tachyon.  We then use
worldsheet methods to study the tachyon condensation in the NSR formulation
of this model, and show that it induces a worldsheet super-Higgs effect.
The main theme of our analysis is the possibility of making meaningful
alternative gauge choices for worldsheet supersymmetry, in place of
the conventional superconformal gauge.  We show in a version of unitary gauge
how the worldsheet gravitino assimilates the goldstino and becomes dynamical.
This picture clarifies recent results of Hellerman and Swanson.  We also
present analogs of $R_\xi$ gauges, and note the importance of logarithmic CFT
in the context of tachyon condensation.}
\begin{document}

\section{Introduction}

The motivation for this paper is to further the studies of time-dependent
backgrounds in string theory.  In particular, we concentrate on the problem
of closed-string tachyon condensation, and its hypothetical relation to the
``spacetime decay to nothing.''

Open-string tachyon condensation is now relatively well-understood 
(see, {\it e.g.}, \cite{sen,headrick} for reviews), as 
a description of D-brane decay into the vacuum (or to lower-dimensional stable
defects).  On the other hand, the problem of the bulk closed-string tachyon
condensation appears related to a much more dramatic instability in which the
spacetime itself decays, or at least undergoes some other extensive change
indicating that the system is far from equilibrium.  In the spacetime
supergravity approximation, this phenomenon has been linked to nonperturbative
instabilities due to the nucleation of ``bubbles of nothing'' \cite{witten}.  
One of the first examples studied in the string and M-theory literature was 
the nonsupersymmetric version of heterotic M-theory \cite{fh}, in which the 
two $E_8$ boundaries of eleven-dimensional spacetime carry opposite relative
orientation and consequently break complementary sets of sixteen
supercharges.  At large separation between the boundaries, this system has an
instanton solution that nucleates ``bubbles of nothing.''  In eleven 
dimensions, the nucleated bubbles are smooth throats connecting the two 
boundaries; the ``nothing'' phase is thus the phase ``on the other side'' 
of the spacetime boundary.  

In addition to this effect, the boundaries are attracted to each other by a 
Casimir force which drives the system to weak string coupling, suggesting some 
weakly coupled heterotic string description in ten dimensions.  In the regime 
of weak string coupling, we expect the originally nonperturbative instability 
of the heterotic M-theory background to turn into a perturbative tachyonic
one.  

We claim that there is a unique viable candidate for describing this
system at weak string coupling:  The tachyonic heterotic string with one
copy of $E_8$ gauge symmetry, and a singlet tachyon.  In this paper, we
study in detail the worldsheet theory of this model -- in the NSR formalism
with local worldsheet $(0,1)$ supersymmetry -- when the tachyon develops a
condensate that grows exponentially along a lightcone direction $X^+$.  There
is a close similarity between this background and the class of backgrounds
studied recently by Hellerman and Swanson \cite{hs1,hs2,hs3,hs4}.%
\footnote{Similar spacetime decay has also been seen in solutions of
noncritical string theory in $1+1$ dimensions \cite{Karczmarek}, and
noncritical M-theory in $2+1$ dimensions \cite{hk1}.}
The main novelty of our approach is
the use of alternative gauge choices for worldsheet supersymmetry, replacing
the traditional superconformal gauge.  We show that the worldsheet dynamics
of spacetime tachyon condensation involves a super-Higgs mechanism, and
its picture simplifies considerably in our alternative gauge.

Our main results were briefly reported in \cite{pixie}; in the present paper,
we elaborate on the conjectured connection to spacetime decay in heterotic
M-theory, and provide more details of the worldsheet theory of tachyon
condensation, including the analysis of the super-Higgs mechanism and its
compatibility with conformal invariance.

Section~2 reviews the nonsupersymmetric version of heterotic M-theory, as a
simple configuration that exhibits the ``spacetime decay to nothing.''  We
argue that the dynamics of this instability should be studied at weak
string coupling, and advocate the role of the tachyonic $E_8$ heterotic
model as a unique candidate for this weakly coupled description of the decay.
In Section~3, we review some of the worldsheet structure of the tachyonic
$E_8$ heterotic string.  In particular, we point out that the $E_8$ current
algebra of the nonsupersymmetric (left-moving) worldsheet sector is realized
at level two and central charge $c_L=31/2$; this is further supplemented by
a single real fermion $\lambda$ of $c_L=1/2$.

Sections~4 and 5 represent the core of the paper, and are in principle
independent of the motivation presented in Section~2.  In Section~4, we
specify the worldsheet theory in the NSR formulation, before and after the 
tachyon condensate is turned on.  The condensate is exponentially growing 
along a spacetime null direction $X^+$.  Conformal invariance then also 
requires a linear dilaton along $X^-$ if we are in ten spacetime dimensions.  
We point out that when the tachyon condensate develops, $\lambda$ transforms 
as a candidate goldstino, suggesting a super-Higgs mechanism in worldsheet 
supergravity.

Section~5 presents a detailed analysis of the worldsheet super-Higgs
mechanism.  Traditionally, worldsheet supersymmetry is fixed by working in
superconformal gauge, in which the worldsheet gravitino is set to zero.  We
discuss the model briefly in superconformal gauge in Section~5.1, mainly to
point out that tachyon condensation leads to logarithmic CFT.  

Since the gravitino is expected to take on a more important role as a result
of the super-Higgs effect, in Section~5.2 and 5.3 we present a gauge choice 
alternative to superconformal.  This alternative gauge choice is inspired by 
the ``unitary gauge'' known from the conventional Higgs mechanism in 
Yang-Mills theories.  We show in this gauge how the worldsheet gravitino 
becomes a dynamical propagating field,
contributing  $c_L=-11$ units of central charge.  Additionally, we analyze
the Faddeev-Popov determinant of this gauge choice, and show that instead
of the conventional right-moving superghosts $\beta$, $\gamma$ of
superconformal gauge, we get {\it left-moving\/} superghosts $\tilde\beta$,
$\tilde\gamma$ of spin 1/2.  In addition, we show how the proper treatment
of the path-integral measure in this gauge induces a shift in the linear
dilaton.  This shift is precisely what is needed for the vanishing of the
central charge when the ghosts are included.  Thus, this string background
is described in our gauge by a worldsheet conformal (but not superconformal)
field theory.  Section~6 points out some interesting features of the
worldsheet theory in the late $X^+$ region, deeply in the condensed phase.

In Appendix~A we list all of our needed worldsheet supergravity conventions.
Appendix~B presents a detailed evaluation of the determinants relevant for
the body of the paper.

\section{Spacetime Decay to Nothing in Heterotic M-Theory}

The anomaly cancelation mechanism that permits the existence of
spacetime boundaries in M-theory works locally near each boundary
component.  The conventional realization, describing the strongly
coupled limit of the $E_8\times E_8$ heterotic string \cite{hw1,hw2}, 
assumes two boundary components, separated by fixed distance $R_{11}$ along 
the eleventh dimension $y$, each breaking the same sixteen supercharges
and leaving the sixteen supersymmetries of the heterotic string.

In \cite{fh}, a nonsupersymmetric variant of heterotic M-theory was
constructed, simply by flipping the orientation of one of the
boundaries. This flipped boundary breaks the complementary set of
sixteen supercharges, leaving no unbroken supersymmetry.  The
motivation behind this construction was to find in M-theory a
natural analog of D-brane anti-D-brane systems whose study turned
out to be so illuminating in superstring theories.  D$p$-branes
differ from $\bar{{\rm D}p}$-branes only in their orientation.  In
analogy with ${\rm D}p$-$\bar{{\rm D}p}$ systems, we refer to the
nonsupersymmetric version of heterotic M-theory as $E_8\times\bar
E_8$ to reflect this similarity.%
\footnote{Actually, this heterotic M-theory configuration is an even closer
analog of a more complicated unstable string theory system: A stack of
D-branes together with an orientifold plane, plus anti-D-branes with an
anti-orientifold plane, such that each of the two collections is separately
neutral.  These collections are only attracted to each other quantum
mechanically, due to the one-loop Casimir effect.}

\subsection{The $E_8\times\bar E_8$ Heterotic M-Theory}

This model, proposed as an M-theory analog of brane-antibrane
systems in \cite{fh}, exhibits two basic instabilities. First, the
Casimir effect produces an attractive force between the two
boundaries, driving the theory towards weak coupling. The strength
of this force per unit boundary area is given by (see \cite{fh} for
details):
\be\CF=-\frac{1}{(R_{11})^{11}}\frac{5}{2^{14}}\int_0^\infty dt\
t^{9/2}\theta_2(0|it),\ee
where $R_{11}$ is the distance between the two branes along the
eleventh dimension $y$.

Secondly, as was first pointed out in \cite{fh}, at large
separations the theory has a nonperturbative instability. This
instanton is given by the Euclidean Schwarzschild solution
\be ds^2=\left(1-\left(\frac{4R_{11}}{\pi
r}\right)^8\right)dy^2+\frac{dr^2}{1-\left(\frac{4R_{11}}{\pi
r}\right)^8}+r^2d^2\Omega_9 \ee
under the ${\bf Z}_2$ orbifold action $y\rightarrow -y$.  Here $r$ and the 
coordinates in the $S_9$ are the other ten dimensions.  
This instanton is schematically depicted in Figure~\ref{figure}(b).

\EPSFIGURE{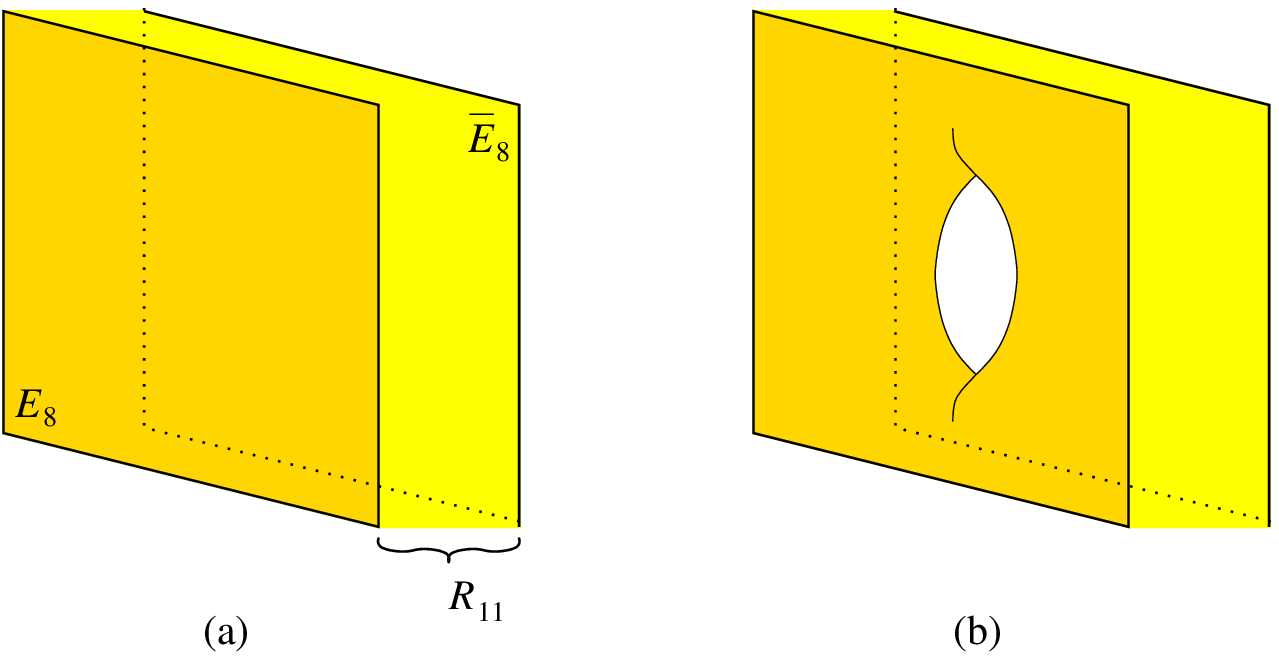}{\label{figure}
(a) A schematic picture of the $E_8\times\bar
E_8$ heterotic M-theory. The two boundaries are separated by
distance $R_{11}$, carry opposite orientations, and support one copy
of $E_8$ gauge symmetry each.  (b)  A schematic picture of the
instanton responsible for the decay of spacetime to ``nothing.'' The
instanton is a smooth throat connecting the two boundaries. Thus,
the ``bubble of nothing'' is in fact a bubble of the hypothetical
phase on the other side of the $E_8$ boundary.}

The probability to nucleate a single ``bubble of nothing'' of this
form is, per unit boundary area per unit time, of order
\be \exp\left(-\frac{4(2R_{11})^8}{3\pi^4G_{10}}\right), \ee
where $G_{10}$ is the ten-dimensional effective Newton constant. As
the boundaries are forced closer together by the Casimir force, the
instanton becomes less and less suppressed. Eventually, there should
be a crossover into a regime where the instability is visible in
perturbation theory, as a string-theory tachyon.

\subsection{The Other Side of the $E_8$ Wall}

The strong-coupling picture of the instanton catalyzing the decay of spacetime 
to nothing suggests an interesting interpretation of this process.  The 
instanton has only one boundary, interpolating smoothly between the two $E_8$ 
walls.  Thus, the bubble of ``nothing'' that is being nucleated 
represents the bubble of a hypothetical phase on the other side 
of the boundary of eleven-dimensional spacetime in heterotic M-theory.  
In the supergravity approximation, this phase truly represents ``nothing,'' 
with no apparent spacetime interpretation.  The boundary conditions at the 
$E_8$ boundary in the supergravity approximation to heterotic M-theory are 
reflective, and the boundary thus represents a perfect mirror.  
However, it is possible that more refined methods, beyond supergravity, may 
reveal a subtle world on the other side of the mirror.  
This world could correspond to a topological phase of the theory, with very 
few degrees of freedom (all of which are invisible in the supergravity 
approximation).

At first glance, it may seem that our limited understanding of M-theory would
restrict our ability to improve on the semiclassical picture of spacetime
decay at strong coupling.  However, attempting to solve this problem at strong
coupling could be asking the wrong question, and a change of perspective
might be in order.  Indeed, the theory itself suggests a less gloomy
resolution: the problem should be properly addressed at weak string coupling,
to which the system is driven by the attractive Casimir force.  Thus, in the
rest of the paper, our intention is to develop worldsheet methods that 
lead to new insight into the hypothetical phase ``behind the mirror,'' in the 
regime of the weak string coupling.

\subsection{Heterotic String Description at Weak Coupling}

We conjecture that when the Casimir force has driven the $E_8$ boundaries into 
the weak coupling regime, the perturbative string description of this system 
is given by the little-studied tachyonic heterotic string model with 
one copy of $E_8$ gauge symmetry \cite{klt}.%
\footnote{Another candidate perturbative description was suggested in
\cite{shanta}.}
The existence of a unique tachyonic $E_8$ heterotic string theory in
ten spacetime dimensions has always been rather puzzling.  We
suspect that its role in describing the weakly coupled stages of the
spacetime decay in heterotic M-theory is the {\it raison d'\^etre\/}
of this previously mysterious model.

We intend to review the structure of this nonsupersymmetric heterotic string 
model in sufficient detail in Section~3.  Anticipating its properties, we list 
some preliminary evidence for this conjecture here:

\begin{itemize}
\item The $E_8$ current algebra is realized at level two.  This is
consistent with the anticipated Higgs mechanism $E_8\times E_8\to E_8$, 
analogous to that observed in brane-antibrane systems where $U(N)\times U(N)$
is first higgsed to the diagonal $U(N)$ subgroup.  (This analogy is discussed 
in more detail in \cite{fh}.)
\item The nonperturbative ``decay to nothing'' instanton instability is
expected to become -- at weak string coupling -- a perturbative instability,
described by a tachyon which is a singlet under the gauge symmetry.  The 
tachyon of the $E_8$ heterotic string is just such a singlet.  
\item The spectrum of massless fermions is nonchiral, with each
chirality of adjoint fermions present.  This is again qualitatively the same
behavior as in brane-antibrane systems.
\item The nonsupersymmetric $E_8\times\bar E_8$ version of heterotic
M-theory can be constructed as a $\Z_2$ orbifold of the standard
supersymmetric $E_8\times E_8$ heterotic M-theory vacuum. Similarly,
the $E_8$ heterotic string is related to the supersymmetric
$E_8\times E_8$ heterotic string by a simple $\Z_2$ orbifold
procedure.
\end{itemize}

The problem of tachyon condensation in the $E_8$ heterotic string
theory is interesting in its own right and can be studied
independently of any possible relation to instabilities in heterotic
M-theory.  Thus, our analysis in the remainder of the paper is
independent of this conjectured relation to spacetime decay in
M-theory.  As we shall see, our detailed investigation of the
tachyon condensation in the heterotic string at weak coupling
provides further corroborating evidence in support of this
conjecture.

\section{The Forgotten $E_8$ Heterotic String}
\label{e8}

Classical Poincar\'e symmetry in ten dimenions restricts the number of
consistent heterotic string theories to nine, of which six are
tachyonic. These tachyonic models form a natural hierarchy,
terminating with the $E_8$ model. We devote this section to a review
of some of the salient aspects of the nearly forgotten heterotic 
$E_8$ theory. Most of these features have been known
for quite some time but are scattered in the literature \cite{klt,dh,forgacs,%
lewellen,elitzur,difrancesco}.

\subsection{The Free Fermion Language}

The tachyonic $E_8$ string was first discovered in the free-fermion
description of the nonsupersymmetric left-movers \cite{klt}.  The
starting point of this construction is the same for all heterotic
models in ten dimensions (including the better-known supersymmetric
models): 32 real left-moving fermions $\lambda^A$, $A=1,\ldots 32$,
and ten right-moving superpartners $\psi_-^\mu$ of $X^\mu$,
described (in conformal gauge; see Appendix~A for our conventions)
by the free-field action
\be S_{\rm fermi}=\frac{i}{2\pi\alpha'}\int d^2\sigma^\pm\,
\left(\lambda_+^A\p_-\lambda_+^A
+\eta_{\mu\nu}\psi_-^\mu\p_+\psi_-^\nu \right). \ee
The only difference between the various models is in the assignment
of spin structures to various groups of fermions, and the consequent
GSO projection. It is convenient to label various periodicity
sectors by a 33-component vector
whose entries take values in $\Z_2=\{\pm\}$,%
\footnote{Here ``$+$'' and ``$-$'' correspond to the NS sector and
the R sector, respectively.  This choice is consistent with the
grading on the operator product algebra of the corresponding
operators.  Hence, the sector labeled by $+$ (or $-$) corresponds to
an antiperiodic (or periodic) fermion on the cylinder.}
\be \BU=(\underbrace{\pm,\ldots\pm}_{32}|\pm). \ee
The first 32 entries indicate the (anti)periodicity of the $A$-th
fermion $\lambda^A$, and the 33rd entry describes the
(anti)periodicity of the right-moving superpartners $\psi^\mu$ of
$X^\mu$.

A specific model is selected by listing all the periodicities that contribute 
to the sum over spin structures.   Modular invariance requires that the
allowed periodicities $\BU$ are given as linear combinations of $n$
linearly independent basis vectors $\BU_i$
\be \BU=\sum_{i=1}^n\alpha_i\BU_i \ee
with $\Z_2$-valued coefficients $\alpha_i$.  Modular invariance also
requires that in any given periodicity sector, the number of
periodic fermions is an integer multiple of eight. All six tachyonic
heterotic theories can be described using the following set of basis
vectors:
\bea \BU_1&=&(--------------------------------|-),\cr
\BU_2&=&(++++++++++++++++----------------|-),\cr
\BU_3&=&(++++++++--------++++++++--------|-),\cr
\BU_4&=&(++++----++++----++++----++++----|-),\cr
\BU_5&=&(++--++--++--++--++--++--++--++--|-),\cr
\BU_6&=&(+-+-+-+-+-+-+-+-+-+-+-+-+-+-+-+-|-).\nonumber \eea
The theory which has only $U_1$ as a basis vector has 32 tachyons.
Adding $U_2$ reduces the number of tachyons to sixteen.  This process can
be continued until the allowed periodicities are spanned by all six
vectors $\BU_i$; here only one bosonic tachyon is present.  This
most intricate of the tachyonic theories is the
single-tachyon $E_8$ model we wish to discuss.%
\footnote{Adding other generators, for example  $\BU_0=(-\ldots
-|+)$ which would relax the lock between the spin structures of the
left and right moving fermions, will produce other heterotic models.
In the case of $\BU_0$ added to any portion of the basis
$\BU_1,\ldots\BU_6$, the familiar $SO(16)\times SO(16)$ model is
produced. Adding generators which have only eight periodic fermions
will produce the remaining supersymmetric theories, as well as extra
copies of the tachyonic ones \cite{klt}.}%

Thus, in our case there are $2^6=64$ different periodicity sectors.
Note that there is a perfect permutation symmetry among the
left-moving fermions, but this symmetry is lifted by the coupling to
the spin structure of the right-moving fermions $\psi^\mu$. There is
precisely one left-moving fermion whose spin structure is always
locked with the spin structure of the supersymmetric sector
$\psi^\mu$.  Since this fermion $\lambda_+^{32}$ plays a special
role, we shall denote it by $\lambda_+$ and refer to it as the
``lone fermion'' for brevity.

The tachyon in this theory is a singlet, and comes from the
$(+\ldots+|+)$ sector in which all the fermions are Neveu-Schwarz.
The left-moving vacuum is excited by the lowest oscillation mode
$b_{-1/2}$ of the lone fermion $\lambda_+$; the right-moving vacuum is in
the ground state.  The vertex operator for the tachyon (in picture 0) is
thus
\be
\CV=(F+\lambda p_\mu \psi^\mu)\exp (i p_\mu X^\mu).
\ee

The spectrum also contains 248 massless vector bosons.  Their spacetime
Yang-Mills group structure is rather obscure in the free fermion language, but
they do form the adjoint of $E_8$.  There is one family of adjoint massless
fermions for each chirality; string loop effects are likely to combine these
into one massive field.  More information about the spectrum at higher levels
can be extracted from the one-loop partition function calculated below, in
(\ref{olpf}).

The 31 free fermions $\lambda^A$, $A=1\ldots 31$ realize a level-two
$E_8$ current algebra.  The $E_8$ current algebra at level $k$ has
central charge
\be
c_{E_8,k}=\frac{k\,{\rm dim}\,E_8}{k+h}=\frac{248k}{k+30}.
\ee
Here $h=30$ is the dual Coxeter number of $E_8$. At level $k=2$,
this corresponds to the central charge of 31/2, which agrees with
the central charge of the 31 free fermions $\lambda^A$ which
comprise it. It is now convenient to switch to a more compact, mixed
representation of the left-moving sector of the worldsheet CFT.\/\
In this representation, the left-movers are succinctly described by
the lone fermion $\lambda_+(\sigma^+)$ together with the algebra of
$E_8$ currents $J^I(\sigma^+)$; here $I$ is the adjoint index of
$E_8$. The spin structure of $\lambda_+$ is locked with the spin
structure of the right-moving superpartners $\psi^\mu_-$ of $X^\mu$.

At level two, the $E_8$ current algebra
has three integrable representations $\ket{\bf 1}$, $\ket{\bf 248}$
and $\ket{\bf 3875}$, where ${\bf n}$ denotes the representation
whose highest weight is in ${\bf n}$ of $E_8$.  The conformal
weights of the highest weight states in $\ket{\bf 1}$, $\ket{\bf
248}$ and $\ket{\bf 3875}$ are 0, 15/16 and 3/2,
respectively.  In the spectrum of physical states, the NS and R sectors
$\ket{\pm}$ of $\lambda_+$ and $\psi_-^\mu$ are interwined with the
representations of $(E_8)_2$, which leads to the following sectors of
the physical Hilbert space,
\bea
\ket{+}&\otimes&\left(\ket{\bf 1}\oplus\ket{\bf 3875}\right),\nonumber\\
\ket{-}&\otimes&\ket{\bf 248}.\nonumber
\eea
One of the advantages of this representation is that the states
charged under the $E_8$ symmetry are now generated by the modes of
the currents $J^I$, making the $E_8$ symmetry manifest and the need
for its realization via 31 free fermions with a complicated GSO
projection obsolete.

\subsection{The Language of Free Bosons}

The bosonization of this model is nontrivial.  In order to rewrite a
free fermion model in terms of bosonic fields, one typically
associates a bosonic field with a pair of fermions.  This is,
however, impossible in the $E_8$ heterotic model: No two fermions
carry the same spin structure in all sectors, due to the intricate
interlacing of the spin structures reviewed above, and a twist of
the conventional bosonization is needed.

As a result, the model cannot be constructed as a straight lattice
compactification; however, it can be constructed as a $\Z_2$
orbifold of one \cite{dh}.  In fact, the starting point can be the
supersymmetric $E_8\times E_8$ heterotic string in the bosonic
language.

The $E_8\times E_8$ lattice of left-moving scalars in the supersymmetric
$E_8\times E_8$ string has a nontrivial outer automorphism $\CI$ that simply
exchanges the two $E_8$ factors.  One can use it to define a $\Z_2$ orbifold
action on the CFT via
\be
\CJ=\CI\cdot\exp\left(\pi iF_s\right),
\ee
where $F_s$ is the spacetime fermion number.  Note that $\exp(\pi iF_s)$ can be
conveniently realized as a $2\pi$ rotation in spacetime, say in the $X^1,X^2$
plane.  The orbifold breaks supersymmetry completely, and yields the
tachyonic $E_8$ heterotic string.

This surprisingly simple orbifold relation between the
supersymmetric $E_8\times E_8$ theory and the tachyonic $E_8$ model
is possible because of some unique properties of the fusion algebra
and the characters of $E_8$.  The fusion rules for the three integrable
highest-weight representations of $(E_8)_2$ are isomorphic to those of a
free CFT of a single real fermion
$\lambda_+$, with $\ket{\bf 248}$ playing the role of the spin
field, and $\ket{\bf 3875}$ that of the fermion.  This is related to
the fact that the $c=1/2$ CFT of $\lambda_+$ can be represented as a
coset
\be
\label{ising}
\frac{(E_8)_1\times(E_8)_1}{(E_8)_2}.
\ee
This explains why there is such a simple relation between the supersymmetric
$E_8\times E_8$ heterotic string and the tachyonic $E_8$ model:  They can be
viewed as two different ways of combining a sigle free-fermion theory 
(\ref{ising}) with the level-two $E_8$ current algebra.

In the bosonic form, the construction of the tachyonic $E_8$
heterotic string model is quite reminiscent of the CHL string
backgrounds  \cite{chl}. In those models, a single copy of $E_8$
symmetry at level two is also obtained by a similar orbifold, but
the vacua are spacetime supersymmetric \cite{cp}.  It is conceivable
that such supersymmetric CHL vacua in lower dimensions could
represent endpoints for decay of the $E_8$ model when the tachyon
profile is allowed an extra dependence on spatial dimensions, as in
\cite{hs1,hs2}.

The one-loop partition function of the heterotic $E_8$ string theory can
be most conveniently calculated in lightcone gauge, by combining the bosonic
picture for the left-movers with the Green-Schwarz representation of the
right-movers.  The one-loop amplitude is given by
\bea
\label{olpf}
{\cal Z}&=&\frac{1}{2}\int\frac{d^2\tau}{({\rm Im}\,\tau)^2}
\frac{1}{({\rm Im}\,\tau)^4|\eta(\tau)|^{24}}\left\{16\Theta_{E_8}(2\tau)
\frac{\theta_{10}^4(0,\bar\tau)}{\theta_{10}^4(0,\tau)}\right.\nonumber\\
& &\qquad\qquad\qquad\left.{}+\Theta_{E_8}(\tau/2)
\frac{\theta_{01}^4(0,\bar\tau)}{\theta_{01}^4(0,\tau)}
+\Theta_{E_8}\left((\tau+1)/2)\right)
\frac{\theta_{00}^4(0,\bar\tau)}{\theta_{00}^4(0,\tau)}\right\}.
\eea
For the remainder of the paper, we use the representation of the worldsheet 
CFT in terms of the lone fermion $\lambda$ and the level-two $E_8$ current 
algebra, represented by 31 free fermions.  

\section{Tachyon Condensation in the $E_8$ Heterotic String}

\subsection{The General Philosophy}

We wish to understand closed-string tachyon condensation as a dynamical
spacetime process.  Hence, we are looking for a time-dependent classical
solution of string theory, which would describe the condensation as it
interpolates between the perturbatively unstable configuration at early times
and the endpoint of the condensation at late times.  Classical solutions of
string theory correspond to worldsheet conformal field theories; thus, in
order to describe the condensation as an on-shell process,  we intend to
maintain exact quantum conformal invariance on the worldsheet.  In particular,
in this paper we are not interested in describing tachyon condensation in
terms of an abstract RG flow between two different CFTs.  In addition, we limit
our attention to classical solutions, and leave the question of string loop
corrections for future work.

\subsection{The Action}

Before any gauge is selected, the $E_8$ heterotic string theory -- with the
tachyon condensate tuned to zero -- is described in the NSR formalism by
the covariant worldsheet action
\bea
\label{covaction}
S_0&=&-\frac{1}{4\pi\alpha'}\int d^2\sigma\,e\left(\vphantom{\frac{1}{2}}
\eta_{\mu\nu}\left(h^{mn}\p_mX^\mu\p_nX^\nu+i\psi^\mu\gamma^m\p_m\psi^\nu
-i\kappa\chi_m\gamma^n\gamma^m\psi^\mu\p_nX^\nu\right)\right.
\nonumber\\
&&\qquad\qquad\qquad\qquad\left.{}+i\lambda^A\gamma^m\p_m\lambda^A
-F^AF^A\vphantom{\frac{1}{2}}
\right).
\eea
where, as usual, $h_{mn}=\eta_{ab}e_m{}^ae_n{}^b$, $e=\det(e_m{}^a)$.
We choose not to integrate out the auxiliary fields $F^A$ from the action at
this stage, thus maintaining its off-shell invariance under local
supersymmetry, whose transformation rules on fields are given by
(\ref{susyright}-\ref{susygrav}).  We have collected other useful formulae
and our choices of conventions in Appendix~\ref{appendix}.

\subsubsection{Linear dilaton}

In order to obtain a description of tachyon condensation in terms of
an exactly solvable CFT, we will consider the tachyon condensate that evolves
along a null direction.  Thus, our tachyon condensate will depend on
a field, say $X^+$, which has trivial OPE with itself. In order for
such a condensate to maintain conformal invariance, we also need to turn
on a linear dilaton background,
\be
\Phi(X)=V_\mu X^\mu,
\ee
for some constant vector $V_\mu$. If we wish to maintain the
critical dimension equal to ten, the linear dilaton must be null,
$V\cdot V=0$.  Hence, we can adjust our choice of spacetime
coordinates $X^{\pm},X^i$ such that $V$ is only nonzero in the $X^-$
direction. Later on, when we turn on the tachyon profile, the linear
dilaton will depend on the light-cone direction $X^+$
instead.%
\footnote{The need for a nonzero dilaton gradient at weak string
coupling is somewhat reminiscent of a similar phenomenon at strong
coupling: the perturbative Casimir force in the $E_8\times\bar E_8$
heterotic M-theory. In both cases, the ``decay-to-nothing''
instability plays out on top of a nontrivial spacetime dependence of
the string coupling.  Given the absence of a reliable interpolation
between the weakly and strongly coupled regimes, it is not possible
to determine whether these two phenomena are directly related.}

In the presence of the linear dilaton, the covariant action of the heterotic
model is $S=S_0+S_V$, with $S_0$ given in (\ref{covaction}) and
\be
\label{dilaction}
S_V=-\frac{1}{4\pi}\int d^2\sigma\,e\,V_\mu\left(\vphantom{M^M}X^\mu R(h)+
i\kappa\chi_{m}\gamma^n\gamma^mD_n\psi^\mu\right).
\ee
Recall that we are in Minkowski signature both on the worldsheet and in
spacetime; this accounts for the negative sign in front of $S_V$. In
the case of the null dilaton, $V\cdot V=0$, both $S_0$ and $S_V$ are
separately Weyl invariant; the proof of this fact for $S_V$ requires
the use of the equations of motion that follow from varying $X^+$ in
the full action.  In addition, off-shell supersymmetry of $S_0+S_V$
also requires a modification of the supersymmetry transformation
rules in the supersymmetric matter sector, which now become
\be
\label{impsusy}
\delta X^\mu=i\epsilon\psi^\mu,\qquad\delta\psi^\mu=\gamma^m\p_m
X^\mu\epsilon+\alpha'V^\mu \gamma^m D_m\epsilon. \ee
The remaining supersymmetry transformations (\ref{susyleft}) and
(\ref{susygrav}) remain unmodified.

The first term in (\ref{dilaction}) produces the standard
$V$-dependent term in the energy-momentum tensor, while the second
term yields the well-known improvement term in the supercurrent of
the linear dilaton theory.  The second term also contributes a
gravitino dependent term to the energy-momentum tensor, as we will 
show below.

\subsubsection{Superpotential and the tachyon profile}

At the classical level, the tachyon couples to the string as a
worldsheet superpotential. Classically, its coupling constant would
be dimensionful. Additionally, the superpotential would be neither
Weyl nor super-Weyl invariant: It would depend on the Liouville mode 
$\phi$ as well as its superpartner $\chi_{-+}$.

We are only interested in adding superpotentials that are, in
conformal gauge, exactly marginal deformations of the original
theory.  The leading-order condition for marginality requires the
tachyon condensate $\CT(X)$ to be a dimension $(1/2,1/2)$ operator,
and the quantum superpotential takes the following form,
\be
\label{qsuppot}
S_W=-\frac{\mu}{\pi\alpha'}\int d^2\sigma\left(\vphantom{\frac{1}{2}}F\,\CT(X)
-i\lambda\psi^\mu\,\p_\mu\CT(X)\right);
\ee
$\mu$ is a dimensionless coupling.

With $\CT(X)\sim\exp(k_\mu X^\mu)$ for some constant $k_\mu$, the condition
for $\CT(X)$ to be of dimension $(1/2,1/2)$ gives
\be
\label{tachonsh}
-k^2+2V\cdot k=\frac{2}{\alpha'}.
\ee
If we wish to maintain quantum conformal invariance at higher orders
in conformal perturbation theory in $\mu$, the profile of the
tachyon must be null, so that $S_W$ stays marginal.  Together with
(\ref{tachonsh}), this leads to
\bea
\label{exptach}
\CT(X)&=&\exp(k_+X^+),\\
\label{onshelltach}
V_-k_+&=&-\frac{1}{2\alpha'}.
\eea
Since our $k_+$ is positive, so that the tachyon condensate grows with
growing $X^+$, this means that $V_-$ is negative, and the theory is weakly
coupled at late $X^-$.

From now on, we will only be interested in the specific form of the
superpotential that follows from (\ref{exptach}) and (\ref{onshelltach}),
\be
\label{qsuppotsp}
S_W=-\frac{\mu}{\pi\alpha'}\int d^2\sigma\left(F-ik_+\lambda_+\psi_-^+\right)
\exp(k_+X^+).
\ee
Interestingly, the check of supersymmetry invariance of (\ref{qsuppotsp})
requires the use of (\ref{onshelltach}) together with the $V$-dependent
supersymmetry transformations (\ref{impsusy}).

\subsection{The Lone Fermion as a Goldstino}

Under supersymmetry, the lone fermion $\lambda_+$ transforms in an interesting
way,
\be
\delta\lambda_+=F\epsilon_+.
\ee
$F$ is an auxiliary field that can be eliminated from the theory by solving
its algebraic equation of motion.  In the absence of the tachyon condensate,
$F$ is zero, leading to the standard (yet slightly imprecise) statement that
$\lambda_+$ is a singlet under supersymmetry.  In our case, when the tachyon
condensate is turned on, $F$ develops a nonzero vacuum expectation value,
and $\lambda_+$ no longer transforms trivially under supersymmetry.  In fact,
the nonlinear behavior of $\lambda_+$ under supersymmetry in the presence of
a nonzero condensate of $F$ is typical of the goldstino.  

Traditionally, the goldstino field $\eta_+$ is normalized such that
its leading order transformation under supersymmetry is just
$\delta\eta_+=\epsilon+\ldots$, where ``$\ldots$'' indicates field-dependent
corrections. In our case, choosing
\be \eta_+=\frac{\lambda_+}{F} \ee
gives the proper normalization for a goldstino under supersymmetry.
Classically, $\eta_+$ transforms as
\be
\delta\eta_+=\epsilon_+-i\eta_+(\epsilon\gamma^mD_m\eta).
\ee
This is the standard nonlinear realization of supersymmetry on the goldstino
in the Volkov-Akulov sense.  This realization of supersymmetry has also played 
a central role in the Berkovits-Vafa construction 
\cite{bv,bastianelli,kunitomo,mca1,mca2}.  This construction has been directly
linked by Hellerman and Swanson to the outcome of tachyon condensation, at
least in the case of Type 0 theory.

\section{Tachyon Condensation and the Worldsheet Super-Higgs Mechanism}

Now that we have precisely defined the worldsheet action in covariant form,
we will show how alternative gauge choices for worldsheet supersymmetry
can elucidate the dynamics of the system, and in particular, make the
worldsheet super Higgs mechanism manifest.

Our alternative gauge choices will have one thing in common with
superconformal gauge:  For fixing the bosonic part of the worldsheet gauge
symmetry, we always pick the conventional conformal gauge, by
setting (locally) $e_m{}^a=\delta_m{}^a$.  This is logically consistent with
the fact that we turn on the tachyon condensate as an exactly marginal
deformation, maintaining worldsheet conformal invariance throughout.
In conformal gauge (and in worldsheet lightcone coordinates
$(\sigma^-,\sigma^+)$), the full worldsheet action becomes
\bea
\label{fulllcaction}
S&=&\frac{1}{\pi\alpha'}\int d^2\sigma^\pm\,\left(\p_+X^i\p_-X^i+\frac{i}{2}
\psi_-^i\p_+\psi_-^i-\p_+X^+\p_-X^--\frac{i}{2}\psi_-^+\p_+\psi_-^-
+\frac{i}{2}\lambda_+^A\p_-\lambda_+^A\right.\nonumber\\
&&\qquad\qquad{}-\mu^2\exp(2k_+X^+)
-i\kappa\chi_{++}\psi_-^i\p_-X^i+\frac{i}{2}\kappa\chi_{++}\psi_-^+\p_-X^-
+\frac{i}{2}\kappa\chi_{++}\psi_-^-\p_-X^+\nonumber\\
&&\qquad\qquad\qquad\qquad\qquad
\left.{}+i\kappa\alpha'V_-\chi_{++}\p_-\psi_-^-
+i\mu k_+\lambda_+\psi_-^+\exp(k_+X^+)\vphantom{\frac{1}{2}}\right).
\eea
In this action, we have integrated out the auxiliaries $F^A$ using their
algebraic equations of motion: All $F^A$s are zero with the exception of the
superpartner $F$ of the lone fermion, which develops a nonzero condensate:
\be
\label{condensate}
F=2\mu\,\exp(k_+X^+).
\ee
The supersymmetry algebra now closes only on-shell, with the use of
the $\lambda_+$ equation of motion.  In the rest of the paper, $F$
always refers to the composite operator in terms of $X^+$ as given
by (\ref{condensate}).  We will use products of powers of $F$ with
other fields several times below. Because the OPE of $F$ with any
field other than $X^-$ is trivial, these objects are quantum
mechanically well-defined so long as $X^-$ does not appear in them.

We also present the energy-momentum tensor and supercurrent, again
in conformal gauge and worldsheet lightcone coordinates:
\bea
T_{++}&=&-\frac{1}{\alpha'}\left\{2i\kappa\chi_{++}\psi^\mu\p_+X_\mu
-2i\kappa\alpha'V_-\left(\frac{3}{2}\chi_{++}\p_+\psi^-_-+\frac{1}{2}(\p_+
\chi_{++})\psi^-_-\right)\right.\nonumber\\
\label{energymomentum}
&&\quad\left.{}+\p_+X^\mu\p_+X_\mu+\frac{i}{2}\lambda^A\p_+
\lambda^A-\alpha'V_-\p_+\p_+X^-\right\},\\
T_{--}&=&-\frac{1}{\alpha'}\left\{\p_-X^\mu\p_-X_\mu+\alpha'V_-\p_-\p_-X^-
+\frac{i}{2}\psi^\mu\p_-\psi_\mu\right\},\\
G_{--}&=&-\frac{2}{\alpha'}\left\{\psi^i_-\p_-X^i-\frac{1}{2}\psi^+_-
\p_-X^--\frac{1}{2}\psi^-_-\p_-X^+-\alpha'V_-\p_-\psi^-_-\right\}.
\label{superc}
\eea
In the classical action (\ref{fulllcaction}), the condensate
(\ref{condensate}) induces a bosonic potential term
${}\sim\mu^2\exp(2k_+X^+)$.  As shown in (\ref{energymomentum}), this
potential term does not contribute to the worldsheet vacuum energy,
since it is an operator of anomalous dimension $(1,1)$ and hence its
integration over the worldsheet does not require any dependence on
the worldsheet metric. Since this potential term does not contribute
to the energy-momentum tensor, it will not contribute to the BRST
charge either.

On the other hand, this bosonic potential does contribute to the equation of
motion for $X^\pm$, which can be written locally as
\bea
\p_+\p_-X^+&=&0,\nonumber\\
\p_+\p_-X^-&=&2\mu^2k_+\exp(2k_+X^+)+\frac{i}{2}\kappa\p_-(\chi_{++}\psi_-^-)
-i\mu k_+^2\lambda_+\psi_-^+\exp(k_+X^+).
\eea
These equations imply that the generic incoming physical excitations of the
string are effectively shielded by the tachyon condensate from traveling too
deeply into the bubble of nothing.  Thus, fundamental string excitations are
pushed away to infinity in the $X^-$ direction by the walls of the ``bubble of
nothing.''  A similar phenomenon has been observed numerous times in the 
previous studies of closed-string tachyon condensation, see {\it e.g.} 
\cite{mcgreevy,horowitz,aharony,hs1,hs2}.

\subsection{Superconformal Gauge}
\label{superconformal}

In the conventional treatment of strings with worldsheet supersymmetry,
superconformal gauge is virtually always selected.  In this gauge, the
worldsheet gravitino is simply set to zero:
\be
\label{scgc}
\chi_{++}=0.
\ee
In our background, however, we expect the gravitino to take on a
prominent role as a result of the super-Higgs mechanism.  For that reason,
we will explore alternative gauge choices, friendlier to this more
important role expected of the gravitino.  

Before we introduce alternative gauge choices, however, we address some 
aspects of the theory in the conventional superconformal gauge.  This exercise 
will reveal at least one intriguing feature of the model:  The emergence
of logarithmic CFT in the context of tachyon condensation.

Superconformal gauge leaves residual superconformal symmetry which
should be realized on all fields by the action of the supercurrent
$G_{--}$. Consider in particular the lone fermion $\lambda_+$.  Before the 
tachyon condensate is turned on, the operator product of $G_{--}$ with 
$\lambda_+$ is nonsingular, in accord with the fact that $\lambda_+$ 
transforms trivially under on-shell supersymmetry.  As we have seen in 
Section~4, when the auxiliary field $F$ develops a nonzero vacuum 
expectation value in the process of tachyon condensation, 
$\lambda_+$ transforms under supersymmetry nontrivially, as a
candidate goldstino.  This raises an interesting question:  How can
this nontrivial transformation be reproduced by the action of
$G_{--}$ on $\lambda_+$, if, as we have seen in (\ref{superc}), the
supercurrent $G_{--}$ is unmodified by $\mu$?

The resolution of this puzzle must come from nontrivial OPEs that develop
at finite $\mu$ between the originally leftmoving field $\lambda_+$ and the
originally rightmoving fields $\psi_-^\pm$.  Here and in the following, it
will be useful to introduce a rescaled version of the fields $\psi_-^\pm$,
\be
\label{firstresc}
\tilde\psi_-^-=\psi_-^-/F,\qquad\tilde\psi_-^+=F\psi_-^+.
\ee
We will encounter this particular rescaled version of $\psi_-^-$
again below, in another gauge.  In terms of $\tilde\psi_-^-$, the
supercurrent (\ref{superc}) simplifies to
\be
\label{supcre}
G_{--}=-\frac{2}{\alpha'}\left\{\psi^i_-\p_-X^i-\frac{1}{2}\psi^+_-\p_-X^-
-\alpha'V_-F\p_-\tilde\psi^-_-\right\}.
\ee
The supersymmetry variations of fields are reproduced as follows.  Consider
for example the supermultiplet $\psi_-^i$ and $X^i$.  In superconformal
gauge, these are free fields, satisfying standard OPEs such as
\be
\psi_-^i(\sigma^\pm)\psi^j(\tau^\pm)
\sim\frac{\alpha'\delta^{ij}}{\sigma^--\tau^-},
\ee
which imply
\be
G_{--}(\sigma^\pm)\psi_-^i(\tau^\pm)\sim\frac{-2\p_-X^i(\tau^\pm)}{
\sigma^--\tau^-},
\ee
correctly reproducing the supersymmetry variation
$\delta\psi_-^i=-2\p_-X^i\epsilon$.

Similarly, the last term in (\ref{supcre}) will reproduce the supersymmetry
transformation $\delta\lambda_+=F\epsilon$ if the following OPE holds,
\be
\label{expectope}
2\alpha'V_-\p_-\psi^-_-(\sigma^\pm)\,\lambda_+(\tau^\pm)\sim
\frac{F(\tau^\pm)/F(\sigma^\pm)}{\sigma^--\tau^-}.
\ee
This required OPE can be checked by an explicit calculation:  Starting with
the free-field OPEs
\be
\psi_-^+(\sigma^\pm)\,\psi_-^-(\tau^\pm)\sim
\frac{-2\alpha'}{\sigma^--\tau^-},
\ee
we get for the rescaled fields (\ref{firstresc})
\bea
\tilde\psi_-^+(\sigma^\pm)\,\tilde\psi_-^-(\tau^\pm)&\sim&
\frac{-2\alpha'F(\sigma^\pm)/F(\tau^\pm)}{\sigma^--\tau^-}\nonumber\\
&&\qquad{}\sim
\frac{-2\alpha'}{\sigma^--\tau^-}
\sum_{n=0}^\infty\frac{1}{n!}(\sigma^+-\tau^+)^n\left(\p_+^n
F(\tau^\pm)\right)/F(\tau^\pm).
\eea
Note that since $X^+$ satisfies locally the free equation of motion
$\p_+\p_- X^+=0$, the coefficient of the $1/(\sigma^--\tau^-)$ term in this
OPE is only a function of $\sigma^+$ and $\tau^+$.
The OPE between $\p_-\tilde\psi_-^-$ and $\lambda_+$ can then be determined
from the the equation of motion for $\lambda_+$:
\be
\p_-\lambda_+(\sigma)=-\mu k_+\psi_-^+\exp(k_+X^+)=-\frac{k_+}{2}
\tilde\psi_-^+.
\ee
Combining these last two equations, we get
\be
\p_-\lambda_+(\sigma^\pm)\tilde\psi_-^-(\tau^\pm)=-\frac{k_+}{2}
\tilde\psi_-^+(\sigma^\pm)\tilde\psi_-^-(\tau^\pm)=\alpha'k_+
\frac{F(\sigma^\pm)/F(\tau^\pm)}{\sigma^--\tau^-}.
\ee
Integrating the result with respect to $\sigma^-$, we finally obtain
\be \label{logope}\lambda_+(\sigma^\pm)\tilde\psi_-^-(\tau^\pm)=
\alpha'k_+\left\{\sum_{n=0}^\infty\frac{1}{n!}(\sigma^+-\tau^+)^n\left(\p_+^n
F(\tau^\pm)\right)/F(\tau^\pm)\right\} \log(\sigma^--\tau^-). \ee
One can then easily check that this OPE implies (\ref{expectope}) when
(\ref{onshelltach}) is invoked.  This in turn leads to
\be
G_{--}(\sigma^\pm)\lambda_+(\tau^\pm)\sim\frac{F(\tau^\pm)}{
\sigma^--\tau^-},
\ee
and the supersymmetry transformation of $\lambda_+$ is correctly reproduced
quantum mechanically, even in the presence of the tachyon condensate.

As is apparent from the form of (\ref{logope}), our theory exhibits --
in superconformal gauge -- OPEs with a logarithmic behavior.  This
establishes an unexpected connection between models of tachyon
condensation and the branch of 2D CFT known as ``logarithmic CFT'' (or LCFT;
see, {\it e.g.}, \cite{flohr,gaberdiel} for reviews).
The subject of LCFT has been vigorously studied in recent years,
with applications to a wide range of physical problems, in particular to
systems with disorder.  The logarithmic behavior of OPEs is
compatible with conformal symmetry, but not with unitarity.  Hence, it
can only emerge in string backgrounds in Minkowski spacetime signature, in
which the time dependence plays an important role (such as our problem of
tachyon condensation).  We expect that the concepts and techniques
developed in LCFTs could be fruitful for understanding time-dependent
backgrounds in string theory.  In the present work, we will not explore this
connection further.

\subsection{Alternatives to Superconformal Gauge}
\label{nunit}

In the presence of a super-Higgs mechanism, another natural choice of
gauge suggests itself.  In this gauge, one anticipates the assimilation of the
Goldstone mode by the gauge field, by simply gauging away the Goldstone mode
altogether.  In the case of the bosonic Higgs mechanism, this gauge is often
referred to as ``unitary gauge'' as it makes unitarity of the theory
manifest.

Following this strategy, we will first try eliminating the goldstino
as a dynamical field, for example by simply choosing
\be
\lambda_+=0
\ee
as our gauge fixing condition for local supersymmetry.  Of course,
this supplements the conformal gauge choice that we have made for
worldsheet diffeomorphisms.  This gauge choice gives rise to
non-propagating bosonic superghosts, together with the usual
propagating fermionic $b,c$ system of central charge $c=-26$.  We
refer to this gauge as ``unitary gauge'' only due to its ancestry in the
similar gauges that proved useful in the Higgs mechanism of Yang-Mills
gauge theories.  In fact, as we shall see, the proper implementation of this
``unitary'' gauge will lead to propagating superghosts, and therefore no
manifest unitarity of the theory.

In addition to the conventional fermionic $b$, $c$ ghosts from
conformal gauge, this gauge choice would lead to non-propagating bosonic
ghosts, which might be integrated out algebraically.   More importantly,
this gauge choice still leaves complicated interaction terms, such as
\be \label{intterm}
\chi_{++}\psi^+_-\p_-X^-,\ee
in the gauge-fixed version of (\ref{fulllcaction}).  Moreover, if the
algebraic superghosts are integrated out, the equation of motion arising
from variation of the action under $\lambda_+$,
\be \mu k_+\psi_-^+\exp(k_+X^+)=0, \ee
needs to be imposed on physical states by hand.  This could be accomplished
by simply setting $\psi_-^+=0$.  This leads to another constraint, imposing
the equation of motion obtained from the variation of $\psi_-^+$ in the
original action as a constraint on physical states.

Instead of resolving such difficulties, we will restart our analysis
with $\psi^+_-=0$ as the gauge condition.  As we will see below,
this condition makes the gauge-fixing procedure transparent.  

\subsection{Liberating the Worldsheet Gravitino in an Alternative Gauge}
\label{liber}

We will now explicitly consider the gauge that begins simply by
setting
\be
\label{pixiegauge}
\psi_-^+=0.
\ee

If $\psi_-^+$ is eliminated from the action, the remaining $\chi_{++}$,
$\psi_-^-$ system can be rewritten as a
purely left-moving first-order system of conformal weights $(3/2,0)$ and
$(-1/2,0)$.%
\footnote{This system is still coupled to the transverse fields $\psi^i$,
$X^i$ but the strength of the interaction goes to zero with growing $F$.}
This can be seen as follows.  Consider first the terms in the
action that are bilinear in these two fields,
\be
\label{bilin}
\frac{1}{2}\chi_{++}\psi_-^-\p_-X^++\alpha'V_-\chi_{++}\p_-\psi_-^-.
\ee
The presence of $\p_-$ here suggests $\chi_{++}$ and $\psi_-^-$
ought to become purely left-moving.  Additionally, these fields show
up only in the energy momentum tensor $T_{++}$ in this gauge.  However, 
their conformal weights do not reflect their left-moving nature. In order 
to obtain fields whose conformal weights are nonzero only in the left-moving
sector, we can rescale
\be
\label{rescale}
\tilde\chi_{++}=F\chi_{++},\qquad\tilde\psi_-^-=\frac{\psi_-^-}{F}.
\ee
This rescaling leads to an additional benefit:  The bilinear terms
(\ref{bilin}) in the classical action now assemble into the canonical kinetic
term of a first-order system of spin 3/2 and $-1/2$,
\be
\label{agra}
\frac{i\kappa V_-}{\pi}\int d^2\sigma^\pm\,\tilde\chi_{++}\p_-
\tilde\psi_-^-,
\ee
with central charge $c_L=-11$.

The first-order system (\ref{agra}) describes the worldsheet gravitino sector
of the theory.  In superconformal gauge in the absence of the tachyon
condensate, the gravitino was non-dynamical and led to a constraint.  Here,
instead, the gravitino has been liberated as a result of the super-Higgs
mechanism: together with its conjugate $\tilde\psi_-^-$, at late
times it appears to have formed a left-moving free-field massless
system, of central charge $c=-11$. In this modified unitary gauge,
the gravitino has been literally set free: It has become a free
propagating massless field!

Our gauge choice (\ref{pixiegauge}) reduces the classical action
significantly, to
\bea
\label{pixieaction}
S&=&\frac{1}{\pi\alpha'}\int
d^2\sigma^\pm\,\left(\p_+X^i\p_-X^i+\frac{i}{2}
\psi_-^i\p_+\psi_-^i-\p_+X^+\p_-X^-
+\frac{i}{2}\lambda_+^A\p_-\lambda_+^A-\mu^2\exp(2k_+X^+)\right.\nonumber\\
&&\qquad\qquad{} \left.{} -i\kappa\chi_{++}\psi_-^i\p_-X^i
+\frac{i}{2}\kappa\chi_{++}\psi_-^-\p_-X^+
+i\kappa\alpha'V_-\chi_{++}\p_-\psi_-^-\vphantom{\frac{1}{2}}
\right). \eea
Note that unlike in the case of superconformal gauge, the gauge fixing
condition (\ref{pixiegauge}) leaves no residual unfixed supersymmetries,
at least at finite $\mu$.  Thus, in this gauge, the theory will be conformal,
but not superconformal.  In Section~\ref{latetimes}, we shall return to the
issue of residual supersymmetry in this gauge, in the late-time limit of
$\mu\to\infty$.

The action (\ref{pixieaction}) will be corrected by one-loop effects.
The first such correction is due to the Faddeev-Popov (super)determinant
$\Delta_{\rm FP}$.
As we will now see, the inherent $X^+$ dependence in our gauge fixing
condition renders $\Delta_{\rm FP}$ dependent on $X^+$ as well.

Our full gauge fixing condition consists of the bosonic conformal
gauge, $e_m{}^a=\delta_m{}^a$, as well as the fermionic condition
(\ref{pixiegauge}).  Note first that the corresponding Faddeev-Popov
superdeterminant factorizes into the ratio of two bosonic determinants,%
\footnote{This is true because the operator whose superdeterminant
is being calculated has a block-triangular form, as the action of
diffeomorphisms on the fermionic gauge-fixing condition $\psi_-^+$
vanishes when (\ref{pixiegauge}) is satisfied.}
\be \label{FadeevPopov}
\Delta_{\rm FP}=J_{bc}/J_{\psi_-^+\epsilon}.
\ee
Here $J_{bc}$ arises from the conformal gauge condition, and
produces the standard set of fermionic $b,c$ ghosts with central
charge $c=-26$. On the other hand, $J_{\psi_-^+\epsilon}$, which
comes from the change of variables between the gauge-fixing
condition $\psi_-^+$ and the infinitesimal supersymmetry parameter
$\epsilon$, turns out to be more complicated.  It will be useful to
rewrite the variation of $\psi^+_-$ as
\bea
\label{op}
\delta\psi^+_-&=& -2\p_-X^+\epsilon+4\alpha'V_-D_-\epsilon\nonumber\\
&=& \frac{4\alpha'V_-}{F}D_-(F\epsilon).
\eea
Here, we are only fixing worldsheet diffeomorphisms and not Weyl
transformations, in order to allow a systematic check of the vanishing of 
the Weyl anomaly.  We are thus in a gauge where 
$h_{mn}=e^{2\phi}\hat{h}_{mn}$. The derivative $D_-$ preserves 
information about the spin of the field it acts on.  We can now see
the denominator of (\ref{FadeevPopov}) should be defined as
\be J_{\psi_-^+\epsilon}=\det\left(\frac{1}{F}D_-F\right). \ee
As in the case of $J_{bc}$, we wish to represent this determinant as
a path integral over a conjugate pair of bosonic fields $\beta$,
$\gamma$: the bosonic ghosts associated with the fixing of local
supersymmetry. Note, however, that the variation of our fixing
condition is dependent on $X^+$. This fact leads to interesting
additional subtleties, which we discuss in detail in
Appendix~\ref{appendixb}.

As a result, the Jacobian turns out to be given by a set of ghosts
$\tilde\beta$, $\tilde\gamma$ of central charge $c=-1$ whose path
integral measure is independent of $X^+$, plus an extra term that depends
explicitly on $X^+$.  As shown in Eqn.~(\ref{Sgauge}) of
Appendix~\ref{appendixb}, we find
\be \label{oneoverJpsi}\frac{1}{J_{\psi_-^+\epsilon}}=
\exp\left\{-\frac{i}{\pi\alpha'}\int
d^2\sigma^\pm\,\alpha'k_+^2\p_+X^+\p_-X^+\right\}
\int\CD\tilde\beta\CD\tilde\gamma\exp\left(iS_{\tilde\beta\tilde
\gamma}\right), \ee
where $S_{\tilde\beta\tilde\gamma}$ is the ghost action for a free 
left-moving bosonic ghost system of spin 1/2.

The Faddeev-Popov determinant thus renormalizes the spacetime metric.  If this
were the whole story, we could re-diagonalize the spacetime metric to the
canonical light-cone form, by redefining the spacetime light-cone coordinates
\bea
Y^+ &=& X^+,\nonumber\\
Y^- &=& X^- +\alpha' k_+^2X^+.
\eea
In these new coordinates, the linear dilaton would acquire a shift:
\bea
\Phi&=&V_-X^-,\nonumber\\
&=& V_-Y^--V_-\alpha'k_+^2X^+,\nonumber\\
&=& V_-Y^- +\frac{k_+}{2}Y^+.
\eea
The effective change in the central charge from this shift in the linear
dilaton would be
\be
c_{dil}=6\alpha'V^2=-24\alpha'V_-V_+=6,
\ee
where we have again used (\ref{tachonsh}).

However, understanding the Faddeev-Popov determinant is not the
whole story. In addition, the $X^+$-dependent rescaling of
$\chi_{++}$ and $\psi_-^-$ as in (\ref{rescale}) also produces a
subtle Jacobian $\tilde J$.  As shown in Appendix~\ref{appendixb},
this Jacobian can be expressed in Minkowski signature as
\be \label{minkjpsi}\tilde J=\exp\left\{-\frac{i}{4\pi\alpha'}\int
d^2\sigma \hat e\left(\alpha'k_+^2\hat
h^{mn}\p_mX^+\p_nX^++\alpha'k_+X^+R\right)\right\}. \ee
In this sense, the original $\chi_{++}$, $\psi_-^-$ system is
equivalent to the canonical $\tilde\chi_{++}$, $\tilde\psi_-^-$
system when this additional renormalization of the linear dilaton
term and the the spacetime metric are taken into account.
In lightcone coordinates, the first factor in (\ref{minkjpsi}) 
becomes 
\be
\exp\left\{\frac{i}{\pi\alpha'}\int d^2\sigma^\pm\,\alpha'k_+^2\p_+X^+
\p_-X^+\right\}.
\ee
This contribution to the renormalization of the metric precisely cancels the 
contribution obtained from the Faddeev-Popov determinant (\ref{oneoverJpsi}).  
Thus, the combined effect of $\tilde J$ and $J_{\psi_-^+\epsilon}$ on the 
$X^\pm$ fields is just a simple shift of the linear dilaton in the original 
$X^\pm$ variables, as implied by the second term in (\ref{minkjpsi}).  
The contribution to the central charge due to this dilaton shift is $c=12$, as 
$c_{dil}=6\alpha'V^2=12$.

The full action, still in conformal gauge, is thus given by:
\bea
\label{cactionn}
S&=&\frac{1}{\pi\alpha'}\int
d^2\sigma^\pm\,\left(\p_+X^i\p_-X^i+\frac{i}{2}
\psi_-^i\p_+\psi_-^i-\p_+X^+\p_-X^-
+\frac{i}{2}\lambda_+^A\p_-\lambda_+^A-\mu^2\exp(2k_+X^+)\right.\nonumber\\
&&\qquad\qquad{} \left.{}
-i\kappa\frac{\tilde\chi_{++}}{F}\psi_-^i\p_-X^i
+i\kappa\alpha'V_-\tilde\chi_{++}\p_-\tilde\psi_-^-\vphantom{\frac{1}{2}}
\right)+S_{bc}+S_{\tilde\beta\tilde\gamma}.
\eea
The information about the linear dilaton profile is absent in this action, but
the energy-momentum tensor of the theory reveals that the dilaton is given by
\be
V_\mu=(k_+,V_-,\vec{0}).
\ee
In this form, the action has only canonical kinetic terms, plus the
potential term and an interaction term which will vanish at late
times (since it goes as $1/F$). Specifically, we note that the fields 
$\tilde\chi_{++}$, $\tilde\psi_-^-$ have now become purely left-moving.  
This matches their conformal weights, which are $(3/2,0)$ and $(1/2,0)$ 
respectively. In addition, they contribute 
only to the left moving energy momentum tensor; their contribution
results in $-11$ units of central charge for $c_L$, in the late $X^+$ limit
when they decouple from the transverse degrees of freedom $X^i,\psi^i$.

Now, let us summarize the central charge contributions of each field
present in this gauge, at late times.  For comparison, we also
present the central charge breakdown for the free theory in 
superconformal gauge.  
\smallskip
\be
\begin{tabular}{cc}
Superconformal gauge & The alternative gauge\\
 {
\begin{tabular}{|l|c|c|}
\hline
Field & $c_L$ & $c_R$ \\
\hline
$X^+$, $X^-$, $X^i$ & $10$ & $10$\\
linear dilaton  & $0$ & $0$\\
$bc$ ghosts &$-26$ & $-26$\\
$\psi^i$ & $0$ & $4$\\
$\psi^+$, $\psi^-$ & $0$ & $1$\\
$\beta$, $\gamma$ ghosts &$0$ & $11$ \\
$\lambda_+$ & $1/2$ & $0$\\
$(E_8)_2$ & $31/2$ & $0$ \\
\hline
\end{tabular}
} & {
\begin{tabular}{|l|c|c|}
\hline
Field & $c_L$ & $c_R$ \\
\hline
$X^+$, $X^-$, $X^i$ & $10$ & $10$\\
linear dilaton & $12$ & $12$\\
$bc$ ghosts &$-26$ & $-26$\\
$\psi^i$ & $0$ & $4$\\
$\tilde\chi$, $\tilde \psi^-$ & $-11$ & $0$ \\
$\tilde\beta$, $\tilde\gamma$ ghosts &$-1$ & $0$ \\
$\lambda_+$ & $1/2$ & $0$\\
$(E_8)_2$ & $31/2$ & $0$ \\
\hline
\end{tabular}
} \end{tabular}
\ee
Our new gauge choice has resulted in $12$ units of central charge in
the right-moving sector, from $\psi^\pm$ and the $\beta\gamma$
system, effectively moving to become $-12$ units of central charge
on the left.  These left-moving central charge units come from
$\tilde\chi$, $\tilde\psi^-$, and the new $\tilde\beta\tilde\gamma$
ghosts. We also see that the shifted linear dilaton precisely
compensates for this relocation of central charge, resulting in the
theory in the alternative gauge still being exactly conformal
at the quantum level.

Interestingly, the equation of motion that follows from varying
$\psi_-^+$ in the original action allows us to make contact with the
original unitary gauge. Classically, this equation of motion is
\be
\label{tildesc}
\p_+\psi_-^-+\kappa\chi_{++}\p_-X^-+2\mu k_+\lambda_+\exp(k_+X^+)=0.
\ee
This constraint can be interpreted and solved in a particularly
natural way: Imagine solving the $X^\pm$ and $\chi_{++}, \psi_-^-$
sectors first.  Then one can simply use the constraint to express
$\lambda_+$ in terms of those other fields. Thus, the alternative 
gauge still allows the gravitino to assimilate the goldstino
in the process of becoming a propagating field.  This is how the 
worldsheet super-Higgs mechanism is implemented, in a way compatible with 
conformal invariance.  

We should note that this classical constraint (\ref{tildesc}) could
undergo a one-loop correction analogous to the one-loop shift in the
dilaton.  We might expect a term $\sim\p_-\chi_{++}$, from varying a
one-loop supercurrent term $\sim\chi_{++}\p_-\psi_-^+$ in the full
quantum action.  Such a correction would not change the fact that
$\lambda_+$ is determined in terms of the oscillators of other fields,
it would simply change the precise details of such a rewriting.

\subsection{$R_\xi$ Gauges}

The history of understanding the Higgs mechanism in Yang-Mills theories
was closely linked with the existence of a very useful family of gauge
choices, known as $R_\xi$ gauges.  $R_\xi$ gauges interpolate -- as one varies
a control parameter $\xi$ -- between unitary gauge and one of the more
traditional gauges (such as Lorentz or Coulomb gauge).

In string theory, one could similarly consider families of gauge fixing
conditions for worldsheet supersymmetry which interpolate between
the traditional superconformal gauge and our alternative gauge.
We wish to maintain conformal invariance of the theory in the new gauge, and
will use conformal gauge to fix the bosonic part of the worldsheet gauge
symmetries.  Because they carry disparate conformal weights $(1,-1/2)$ and 
$(0,1/2)$ respectively, we cannot simply add the two gauge fixing conditions 
$\chi_{++}$ and $\psi_-^+$ with just a relative constant.   
In order to find a mixed gauge-fixing condition compatible with conformal
invariance, we need a conversion factor that makes up for this difference in
conformal weights.  One could for example set
\be
\label{mixed}
(\p_-X^+)^2\chi_{++}+\xi F^2\psi_-^+=0,
\ee
with $\xi$ a real constant (of conformal dimension 0).  The added
advantage of such a mixed gauge is that it interpolates between
superconformal and alternative gauge not only as one changes $\xi$,
but also at any fixed $\xi$ as $X^+$ changes:  At early lightcone
time $X^+$, the superconformal gauge fixing condition dominates,
while at late $X^+$, the alternative gauge takes over, due to the
relative factors of $F$ between the two terms in (\ref{mixed}).

Even though (\ref{mixed}) is compatible with conformal invariance, it is
highly nonlinear, and therefore impractical as a useful gauge fixing
condition.  Another, perhaps more practical, condition that incorporates
both $\chi_{++}$ and $\psi_-^+$ is
\be
\label{interesting}
\p_-\chi_{++}+\xi\p_+\psi_-^+=0.
\ee
Here the mismatch in conformal weights has been made up by inserting
worldsheet derivatives, rather than other composite operators.  This
condition can be written in the following covariant form,
\be
\gamma^m\gamma^nD_n\chi_m-2\xi\gamma^mD_m\psi^+=0,
\ee
demonstrating its compatibility with diffeomorphism invariance.

Note that
(\ref{interesting}) correctly anticipates the dynamics of the gravitino:
If we solve for $\p_+\psi_-^+$ using (\ref{interesting}) in the kinetic term
$\psi_-^-\p_+\psi_-^+$, we get $\psi_-^-\p_-\chi_{++}$.  Thus, $\chi_{++}$
replaces $\psi_-^+$ as the conjugate partner of $\psi_-^-$, and turns it from
a right-moving field into a left-moving one.  Note also that both terms in
(\ref{interesting}) transform under supersymmetry as
$\sim\p_+\p_-\epsilon+\ldots $, which implies that the bosonic superghosts
$\hat\beta$, $\hat\gamma$ associated with this gauge fixing will have a
second-order kinetic term,
\be
\sim\int d^2\sigma^\pm\,\left(\hat\beta\p_+\p_-\hat\gamma+\ldots\right).
\ee
In the $\mu\to\infty$ limit, this gauge can be expected to leave a residual 
fermionic gauge symmetry.  

Various other classes of $R_\xi$-type gauges can be considered.
For example, one could study combinations of superconformal gauge with
the naive unitary gauge in which $\lambda_+$ is set to zero, leading to
gauge fixing conditions such as
\be
F\chi_{++}+\xi k_+\lambda_+\p_+X^+=0,
\ee
or
\be
F\lambda_++\xi\chi_{++}k_+\p_-X^+=0.
\ee
Another interesting possibility is
\be
\chi_{++}+\xi\,\p_+\!\left(\frac{\lambda_+}{F}\right)=0,
\ee
which would interpolate between superconformal gauge and a weaker form of
unitary gauge, in which the left-moving part of the goldstino is set to zero.

We have not investigated the worldsheet theory in these mixed gauges, but it
would be interesting to see if any of them shed some new light on the
dynamics of tachyon condensation.

\section{The Condensed Phase: Exploring the Worldsheet Theory at $\mu=\infty$}
\label{latetimes}

The focus of the present paper has been on developing worldsheet
techniques that can elucidate the super-Higgs mechanism and the
dynamics of the worldsheet gravitino in the process of spacetime
tachyon condensation. Here we comment briefly on the structure of
the worldsheet theory in the regime where the tachyon has already
condensed.

This condensed phase corresponds to the system at late $X^+$.  In the
worldsheet theory, a constant translation of $X^+$ rescales the value of
the superpotential coupling $\mu$, with the late $X^+$ limit mapping to
$\mu\to\infty$.   In that limit, the worldsheet theory simplifies in an
interesting way.  First, we rescale the parameter of local supersymmetry 
transformation,
\be
\tilde\epsilon_+=F\epsilon_+.
\ee
The supersymmetry variations then reduce in the $\mu=\infty$ limit and 
in conformal gauge to
\bea
\label{latesusy}
\delta X^+&=&0,\qquad\qquad\qquad\delta\tilde\psi_-^+=4\alpha'V_-\p_-
\tilde\epsilon_+,\nonumber\\
\delta X^-&=&-i\tilde\psi_-^-\tilde\epsilon_+,\quad\qquad\delta
\tilde\psi_-^-=0,\\
\delta X^i&=&0,\qquad\qquad\qquad\delta\tilde\psi_-^i=0,\nonumber\\
\delta\lambda_+&=&\tilde\epsilon_+,\qquad\qquad\quad
\delta\tilde\chi_{++}=\frac{2}{\kappa}\left(\p_+\tilde\epsilon_+-k_+\p_+X^+
\tilde\epsilon_+\right).\nonumber
\eea
Note that $X^+$ is now invariant under supersymmetry.  Consequently, the
terms in the action that originate from the superpotential,
\be
-2\mu^2\exp(2k_+X^+)-ik_+\lambda_+\tilde\psi_-^+,
\ee
are now separately invariant under (\ref{latesusy}), in the strict
$\mu=\infty$ limit.  This in turn implies that we are free to drop the
potential term $\sim\mu^2\exp(2k_+X^+)$ without violating supersymmetry.
The resulting model is then described, in the alternative gauge of
Section~\ref{liber}, by a free field action:
\bea
S_{\mu=\infty}&=&\frac{1}{\pi\alpha'}\int
d^2\sigma^\pm\,\left(\p_+X^i\p_-X^i+\frac{i}{2}
\psi_-^i\p_+\psi_-^i-\p_+X^+\p_-X^-
+\frac{i}{2}\lambda_+^A\p_-\lambda_+^A\right.\nonumber\\
&&\qquad\qquad\qquad{} \left.{}
+i\kappa\alpha'V_-\tilde\chi_{++}\p_-\tilde\psi_-^-\vphantom{\frac{1}{2}}
\right)+S_{bc}+S_{\tilde\beta\tilde\gamma}.
\eea

We argued in Section~\ref{liber} that at finite $\mu$, our
alternative gauge (\ref{pixiegauge}) does not leave any residual
unfixed supersymmetry, making the theory conformal but not
superconformal.  This is to be contrasted with superconformal gauge,
which leaves residual right-moving superconformal symmetry.  At
$\mu=\infty$, however, it turns out that the alternative gauge
(\ref{pixiegauge}) does leave an exotic form of residual supersymmetry.
Note first that at any $\mu$, the fermionic gauge fixing condition is
respected by $\tilde\epsilon$ that satisfy
\be
\label{ress}
\p_-\epsilon+k_+\p_-X^+\epsilon\equiv\frac{1}{F}\p_-\tilde\epsilon_+=0.
\ee
At finite $\mu$, these apparent residual transformations do not preserve the
bosonic part of our gauge fixing condition, $e_m{}^a=\delta_m^a$.
In the $\mu=\infty$ limit, however, the supersymmetry transformation of
$e_m^a$ is trivial, and all solutions of (\ref{ress}) survive as 
residual supersymmetry transformations.

Unlike in superconformal gauge, this residual supersymmetry is
{\it left-moving\/}.  
Moreover, the generator $\tilde\epsilon_+$ of local supersymmetry is of
conformal dimension $(1/2,0)$; hence, local supersymmetry cannot be expected
to square to a worldsheet conformal transformation.  As is clear from
(\ref{latesusy}), this local supersymmetry is in fact nilpotent.

This residual supersymmetry $\tilde\epsilon(\sigma^+)$ can be fixed
by a supplemental gauge choice, setting a purely left-moving fermion
to zero.  The one that suggests itself is $\lambda_+$, which in this
gauge satisfies the free equation of motion; moreover, $\lambda_+$
transforms very simply under $\tilde\epsilon$ supersymmetry, and
setting it to zero fixes that symmetry completely.  Once we add the
condition $\lambda_+=0$ to the gauge choice $\psi_-^+=0$, this gauge
now becomes very similar to the naive unitary gauge of
Section~\ref{nunit}.  This is another way of seeing that the
goldstino is not an independent dynamical field, since it has been
absorbed into the dynamics of the other fields in the process of
setting the gravitino free.

Alternatively, one can leave the residual supersymmetry unfixed, and instead
impose the constraint (\ref{tildesc}) which reduces in the $\mu=\infty$ limit
to
\be
\lambda_+=2\alpha'V_-\p_+\tilde\psi_-^-.
\ee

Depending on whether the coupling of the quantum bosonic potential
$\exp(2k_+X^+)$ is tuned to zero or not, we have two different
late-time theories:  One whose equations of motion expel
all degrees of freedom to future infinity along $X^-$ for some
constant $X^+$, the other allowing perturbations to reach large
$X^+$. It is the latter theory which may represent a good set of variables
suitable for understanding the physics at late $X^+$. If the potential is
retained, the physical string modes are pushed away to infinity along $X^-$
before reaching too deeply into the condensed phase, confirming that very few 
of the original stringy degrees of freedom are supported there.  

\section{Conclusions}

\subsection{Overview}

The main focus of this paper has been on examining the worldsheet theory 
of tachyon condensation in the $E_8$ heterotic string model.  We studied 
the theory with a linear dilaton $V=V_-X^-$, and a tachyon profile 
$\CT(X^\mu)=2\mu\exp(k_+X^+)$.  At first glance, this tachyon profile
produces a worldsheet potential term which expels all degrees of
freedom from the large $X^+$ region, indicating a possible
topological phase, conjecturally related to the ``nothing'' phase in 
heterotic M-theory.  

We found that the worldsheet dynamics of tachyon condensation involves a 
super-Higgs mechanism, and that its analysis simplifies when local 
worldsheet supersymmetry is fixed in a new gauge, specifically conformal gauge
augmented by $\psi^+_-=0$.  Following a detailed analysis of
one-loop measure effects, we found that exact quantum conformal invariance is
maintained throughout in this gauge.  At late times, the worldsheet theory
contains a free left-moving propagating gravitino sector 
(obscured in superconformal gauge, as there the gravitino is set to zero).  
The gravitino sector contributes $-11$ units of central charge to the 
left-movers.  In addition, the gauge fixing leads to a set of left-moving 
ghosts with $c=-1$, and a spacelike shifted linear dilaton $V=V_-X^-+k_+X^+$.

In the process of making the gravitino dynamical, the worldsheet goldstino 
$\lambda_+$ has been effectively absorbed into the rest of the system; 
more precisely, the constraint generated by the alternative gauge can be 
solved by expressing $\lambda_+$ in terms of the remaining dynamical degrees 
of freedom.  

\subsection{Further Analysis of the $E_8$ System}

In this paper, we have laid the groundwork for an in-depth analysis
of the late time physics of the $E_8$ heterotic string under tachyon
condensation.  The emphasis here has been on developing the worldsheet 
techniques, aimed in particular at clarifying the super-Higgs mechanism.  
The next step, which we leave for future work, would be to examine the 
spacetime physics in the regime of late $X^+$ where the tachyon has 
condensed.  There are signs indicating that this phase contains very few 
conventional degrees of freedom;  more work is needed to provide further 
evidence for the conjectured relation between tachyon condensation in the 
$E_8$ string and the spacetime decay to nothing in $E_8\times\bar E_8$ 
heterotic M-theory.

It would be interesting to use the standard tools of string theory, combined 
with the new worldsheet gauge, to study the spectrum and scattering amplitudes 
of BRST invariant states in this background, in particular at late times.  The 
use of mixed $R_\xi$ gauges could possibly extend the range of such an 
analysis, by interpolating between the superconformal gauge and its 
alternative.

\subsection{Towards Non-Equilibrium String Theory}

In the process of the worldsheet analysis presented in this paper, we 
found two features which we believe may be of interest to a broader 
class of time-dependent systems in string theory: 
(1) in superconformal gauge, the spacetime tachyon condensate turns the 
worldsheet theory into a logarithmic CFT; and 
(2) the worldsheet dynamics of some backgrounds may simplify 
in alternative gauge choices for worldsheet supersymmetry. 

We have only explored the first hints of the LCFT story and its utility in the 
description of string solutions with substantial time dependence.  
The new gauge choices, however, are clearly applicable to other systems.  
As an example, consider the Type 0 model studied 
in \cite{hs3}.  We can pick a gauge similar to our alternative gauge 
(\ref{pixiegauge}) by setting $\psi^+_-=\psi^+_+=0$, again in addition to
conformal gauge.  We expect to simply double the gauge fixing
procedure in Section \ref{liber}, producing one copy of $c=-1$ superghosts 
and one copy of the propagating gravitino sector in both the left and right 
movers.  When the one-loop determinant effects are included, the linear 
dilaton shifts by $2k_+$, resulting in additional 24 units of central charge.  
Together with the two $c=-11$ sectors and the two $c=-1$ ghost sectors, this 
shift again leads $c_{tot}=0$, similarly to the heterotic model studied in 
the present paper.  Exact conformal invariance is again 
maintained at the quantum level.  This is in accord with the results of
\cite{hs3}; however, our results are not manifestly equivalent to those of 
\cite{hs3}, as a result of a different gauge choice.  In addition, our results 
also suggest an interpretation of the somewhat surprising appearance of 
fermionic first-order systems with $c=-11$ in \cite{hs3}:  It is likely that 
they represent the re-emergence of the dynamical worldsheet gravitino, in 
superconformal gauge.  

This picture suggest a new kind of worldsheet duality in string theory.  
Instead of viewing one CFT in two dual ways while in superconformal gauge, 
the new duality is between two different CFTs representing the same solution 
of string theory, but in two different worldsheet gauges.  The physics of 
BRST observables should of course be gauge independent, but this does not 
require the CFTs to be isomorphic.  In fact, the alternative gauge studied 
in the body of this paper represents an example: It leads to a conformal, 
but not superconformal theory, yet it should contain the same physical 
information as the SCFT realization of the same string background in 
superconformal gauge.   

We hope that this technique of using gauge choices other than superconformal 
gauge will be applicable to a wider class of time dependent string 
backgrounds, beyond the case of models with tachyon condensation studied here, 
and that it will increase our understanding of non-equilibrium string 
theory.

\acknowledgments

One of us (PH) wishes to thank Allan Adams, Michal Fabinger, Uday Varadarajan
and Bruno Zumino for useful discussions in the early stages of this project, 
in 2001-02.  This work has been supported by NSF Grants PHY-0244900 and
PHY-0555662, DOE Grant DE-AC03-76SF00098, an NSF Graduate Research Fellowship,
and the Berkeley Center for Theoretical Physics.

\appendix
\section{Supersymmetry Conventions}
\label{appendix}

In the NSR formalism, the tachyonic $E_8$ heterotic string is described by
worldsheet supergravity with $(0,1)$ supersymmetry.%
\footnote{A useful source of information on two-dimensional supergravities is,
{\it e.g.}, \cite{orange}.}
Here we list our conventions for worldsheet supergravity.

\subsection{Flat worldsheet}

In the worldsheet coordinates $\sigma^a=(\sigma^0\equiv\tau,\sigma^1
\equiv\sigma)$, the flat Lorentz metric is
\begin{equation}
\eta_{ab}=\pmatrix{-1&0\cr 0&1},
\end{equation}
with gamma matrices given by
\begin{equation}
\gamma^0=\pmatrix{0&1\cr -1&0},\qquad
\gamma^1=\pmatrix{0&1\cr 1&0},
\end{equation}
and satisfying $\{\gamma^a,\gamma^b\}=2\eta^{ab}$.  We define the chirality
matrix
\be
\Gamma\equiv\gamma^0\gamma^1=\pmatrix{1&0\cr 0&-1}.
\ee

The two-component spinor indices $\alpha$ are raised and lowered using the
natural symplectic structure,
\be
\xi^\alpha\equiv\varepsilon^{\alpha\beta}\xi_\beta,\qquad
\varepsilon^{\alpha\beta}=\pmatrix{0&-1\cr 1&0},
\ee
with $\varepsilon^{\alpha\beta}\varepsilon_{\beta\gamma}=
\varepsilon_{\gamma\beta}\varepsilon^{\beta\alpha}=\delta^\alpha_\gamma$.
This implies, for example, that $\xi^\alpha\zeta_\alpha=\zeta^\alpha\xi_\alpha=
-\xi_\alpha\zeta^\alpha$ and $\xi\gamma^a\zeta=-\zeta\gamma^a\xi$, for any
two real spinors $\xi,\zeta$.

Any two-component spinor $\xi_\alpha$ can be decomposed into its chiral
components, defined via
\be
\xi_\alpha\equiv\pmatrix{\xi_+\cr\xi_-},\qquad
\xi_\pm=\frac{1}{2}(1\pm\Gamma)\xi.
\ee

\subsection{Local worldsheet supersymmetry}

On curved worldsheets, we will distinguish the worldsheet index $m$
from the internal Lorentz index $a$.  The spacetime index $\mu$ runs
over $0,\ldots D-1$, typically with $D=10$. The heterotic string
action with $(0,1)$ worldsheet supersymmetry is given by
\bea
\label{hetaction}
S&=&-\frac{1}{4\pi\alpha'}\int d^2\sigma\,e\left(\vphantom{\frac{1}{2}}
\eta_{\mu\nu}\left(h^{mn}\p_mX^\mu\p_nX^\nu+i\psi^\mu\gamma^m\p_m\psi^\nu
-i\kappa\chi_m\gamma^n\gamma^m\psi^\mu\p_nX^\nu\right)\right.
\nonumber\\
&&\qquad\qquad\qquad\qquad\left.{}+i\lambda^A\gamma^m\p_m\lambda^A
-F^AF^A\vphantom{\frac{1}{2}} \right). \eea
The fermions and gravitino satisfy the following chirality
conditions:
\be \Gamma \psi^\mu_-=-\psi^\mu_-, \qquad \Gamma
\lambda_+=\lambda_+, \qquad \Gamma \chi_+=\chi_+ \ee
The action (\ref{hetaction}) is invariant under the supersymmetry
transformations given by
\be
\label{susyright}
\delta X^\mu=i\epsilon\psi^\mu,\qquad\delta\psi^\mu=\gamma^m\p_m
X^\mu\epsilon
\ee
for the right-moving sector,
\be
\label{susyleft}
\delta\lambda^A=F^A\epsilon,\qquad\ \delta F^A=i\epsilon\gamma^mD_m\lambda^A
\ee
for the left-movers, and
\be
\label{susygrav}
\delta e_m{}^a=i\kappa\epsilon\gamma^a\chi_m,\qquad
\delta\chi_m=\frac{2}{\kappa}D_m\epsilon
\ee
in the supergravity sector. Note that we set $\kappa=2$ in \cite{pixie}.  
Of course, $\gamma^m=e^m{}_a\gamma^a$, and the covariant derivative on 
spinors is
\be
D_m\zeta=\left(\p_m+\frac{1}{4}\omega_m{}^{ab}\gamma_{ab}\right)\zeta,
\ee
with $\gamma^{ab}=[\gamma^a,\gamma^b]/2$.  In general, the spin connection
$\omega_m{}^{ab}$ in supergravity contains the piece that depends solely on
the vielbein, $\omega_m{}^{ab}(e)$, plus a fermion bilinear improvement term.
In conformal $(0,1)$ supergravity in two dimensions as described by
(\ref{hetaction}), however, the improvement term vanishes identically, and
we have $\omega_m{}^{ab}=\omega_m{}^{ab}(e)$, with
\be
\omega_m{}^{ab}=\frac{1}{2}e^{na}(\p_m e_n{}^b-\p_n e_m{}^b)
-\frac{1}{2}e^{nb}(\p_m e_n{}^a-\p_n e_m{}^a)
-\frac{1}{2}e^{na}e^{pb}(\p_n e_{pc}-\p_p e_{nc})e_m{}^c.
\ee

Note also that the susy variation of $F^A$ is sometimes written in the
literature as $\delta F^A=i\epsilon\gamma^m\hat D_m\lambda^A$, using the
supercovariant derivative
\be
\label{gross}
\hat D_m\lambda^A_+\equiv\left(\p_m+\frac{1}{4}\omega_m{}^{ab}\gamma_{ab}
\right)\lambda_+^A-\chi_{m+}F^A.
\ee
This simplifies, however, in several ways.  First of all, the gravitino drops
out from (\ref{gross}) if the $(0,1)$ theory is independent of the
superpartner of the Liouville field, as is the case for our heterotic
worldsheet supergravity.  Secondly, for terms relevant for the action, we
get
\be
\label{lovely}
\lambda^A\gamma^m\hat D_m\lambda^A_+\equiv\lambda^A\gamma^m\p_m\lambda^A_+.
\ee

\subsection{Lightcone coordinates}

The worldsheet lightcone coordinates are
\be
\sigma^\pm=\tau\pm\sigma,
\ee
in which the Minkowski metric becomes
\be
\eta_{ab}=\pmatrix{0&-\frac{1}{2}\cr -\frac{1}{2}&0},
\ee
resulting in the lightcone gamma matrices
\be
\gamma^+=\pmatrix{0&2\cr 0&0},\qquad
\gamma^-=\pmatrix{0&0\cr -2&0}.
\ee
On spin-vectors, such as the gravitino $\chi_{m\alpha}$, we put the
worldsheet index first, and the spinor index second when required.
We will use $\pm$ labels for spinor indices as well as lightcone
worldsheet and spacetime indices, as appropriate.  The nature of a
given index should be clear from context. In lightcone coordinates,
and in conformal gauge, the supersymmetry transformations of the
matter multiplets are given by
\bea
\delta X^\mu&=&i\epsilon_+\psi^\mu_-,\qquad\delta\psi_-=-2\p_-X^\mu\epsilon_+,
\nonumber\\
\delta\lambda_+^A&=&F^A\epsilon_+,\qquad\ \delta F^A=-2i\epsilon_+
\p_-\lambda_+^A,
\eea
and the linearized supersymmetry transformations of the supergravity
multiplet are
\be
\delta e_+{}_-=-2i\kappa\epsilon_+\chi_{++},\qquad\delta\chi_{++}=
\frac{2}{\kappa}\p_+\epsilon_+.
\ee

Once we have picked conformal gauge, we can meaningfully assign a
conformal weight to each field.  The chart below lists these
conformal dimensions for all relevant objects:
\smallskip
\be
\begin{tabular}{|l|c|l|}
\hline
Field & Symbol & Conformal Weight \\
\hline
gravitino & $\chi_{++}$  & $(1,-\frac{1}{2})$ \\
goldstino & $\lambda_+$    & $(\frac{1}{2},0)$ \\
fermion   & $\psi^\mu_-$ & $(0,\frac{1}{2})$ \\
boson     & $X^\mu$      & $(0,0)$            \\
aux.\ field & $F$          & $(\frac{1}{2},\frac{1}{2})$ \\
SUSY\ parameter & $\epsilon$ & $(0,-\frac{1}{2})$ \\
w.s.\ derivative& $\p_-$ & $(0,1)$           \\
w.s.\ derivative& $\p_+$ & $(1,0)$            \\
\hline
\end{tabular}
\ee

\noindent In spacetime, similarly, we define the lightcone
coordinates thus:
\be
X^\pm=X^0\pm X^1,
\ee
and denote the remaining transverse dimensions by $X^i$, so that the spacetime
index decomposes as $\mu\equiv(+,-,i)$.

When we combine the spacetime lightcone parametrization with the lightcone
coordinate choice on the worldsheet, the heterotic action becomes
\bea
\label{lcaction}
S&=&\frac{1}{\pi\alpha'}\int d^2\sigma^\pm\,\left(\p_+X^i\p_-X^i+\frac{i}{2}
\psi_-^i\p_+\psi_-^i-\frac{1}{2}\p_+X^+\p_-X^--\frac{1}{2}\p_+X^-\p_-X^+\right.
\nonumber\\
&&\qquad\qquad\qquad\qquad-\frac{i}{4}\psi_-^+\p_+\psi_-^--\frac{i}{4}
\psi_-^-\p_+\psi_-^++\frac{i}{2}\lambda_+^A\p_-\lambda_+^A+\frac{1}{4}F^AF^A\\
&&\qquad\qquad\qquad\qquad\qquad\left.{}-\frac{i}{2}\kappa\chi_{++}
(2\psi_-^i\p_-X^i-\psi_-^+\p_-X^--\psi_-^-\p_-X^+)\right),\nonumber
\eea
where we have defined $d^2\sigma^\pm\equiv d\sigma^-\wedge d\sigma^+
=2d\tau\wedge d\sigma$.

\section{Evaluation of the Determinants}
\label{appendixb}

We wish to calculate the determinant for the operator
\be D_F\equiv \frac{1}{F}D_-F \ee
as it acts on fields of arbitrary half-integer spin $j$. $F$ here is
a shorthand for the tachyon condensate $F=2\mu\exp(k_+X^+)$, as
determined in (\ref{condensate}); we will continue with this
notation through this appendix. As $D_F$ is a chiral operator, we
will take the usual approach of finding its adjoint and calculating
the determinant of the corresponding Laplacian. The actual
contribution we are interested in will be the square root of the
determinant of the Laplacian.  Throughout this appendix, we work in
Euclidean signature; we will Wick rotate our results back to
Minkowski signature before adding the results to the body of the
paper.  Also, we gauge fix only worldsheet diffeomorphisms by
setting
\be h_{mn}=e^{2\phi}\hat{h}_{mn},\ee
where $\phi$ is the Liouville field.  For most calculations, we set
the fiducial metric $\hat{h}_{mn}$ to be the flat metric. In this
gauge, we find
\be D_-=e^{-2\phi}\bar\p, \ee
independently of $j$.

We define the adjoint $D_F^\dagger$ of the Faddeev-Popov operator
$D_F$ via
\be
\label{conjdef}
\bra{T_1}\frac{1}{F}D_-(FT_2)\rangle=\bra{D_F^\dagger T_1}T_2\rangle,
\ee
where $T_1$ is a worldsheet tensor of spin $j-1$, $T_2$ is a tensor
of spin $j$, and the inner product on the corresponding tensors is
the standard
one, independent of $F$.%
\footnote{See Section II.E of \cite{dhoker} for more details on the
corresponding inner products and the definition and properties of
differential operators on Riemann surfaces.}

We find the left hand side of (\ref{conjdef}) becomes
\bea
\bra{T_1}\frac{1}{F}D_-FT_2\rangle=\int d^2z
e^{2\phi(2-j)}T_1^*\left(\frac{1}{F}D_-F\right)T_2&=&\int d^2z
e^{2\phi(2-j)}T_1^*\frac{1}{F}e^{-2\phi}\bar\p\left(FT_2\right)\nonumber\\
{}=-\int d^2z \bar\p\left(e^{-2\phi
(j-1)}\frac{1}{F}T_1^*\right)FT_2 = &-&\!\int d^2z e^{2\phi
(1-j)}\left(F D_+ \frac{1}{F}T_1\right)^{\!*}\!T_2.
\eea
Thus, the adjoint operator is
\be D_F^\dagger=-F D_+ \frac{1}{F}, \ee
where
\be
D_+=e^{2\phi(j-1)}\p e^{-2\phi(j-1)} \ee
when acting on a field of spin $j-1$. As above, our conventions are
such that $D_+$ and $\p$ act on everything to their right.

We are now interested in the determinant of the Laplace operator
$D_F^\dagger D_F$,
\be \det
\left(-\frac{1}{F}D_+F^2D_-\frac{1}{F}\right)=\det\left(-\frac{1}{F^2}D_+F^2D_-
\right). \ee
Determinants of such operators were carefully evaluated
in \cite{Kallosh};%
\footnote{Similar determinants have played a central role in other areas of
CFT, perhaps most notably in the free-field Wakimoto realizations of WZW
models; see, {\it e.g.}, \cite{gerasimov}.}
Eqn.~(3.2) of that paper gives a general formula for the
determinant of $f\p g\bar{\p}$. For our case,
\be f=e^{(2j-2)\phi+2k_+X^+},\qquad g=e^{-2j\phi-2k_+X^+},
\ee
and we get
\bea \label{trivialdet} \log\det\left(D_F^\dagger D_F\right)
&=&-\frac{1}{24\pi}\int d^2\sigma\, \hat
e\left\{[3(2j-1)^2-1]\hat{h}^{mn}\p_m\phi\p_n\phi
\right.\nonumber\\
&+&12(2j-1)k_+\hat{h}^{mn}\p_m\phi\p_n X^+\left.+
12k_+^2\hat{h}^{mn}\p_m X^+\p_n X^+\right\}. \eea

\subsection{$J_{\psi_-^+\epsilon}$}

The case of the Faddeev-Popov operator corresponds to spin $j=1/2$.
Since we are in fact interested in the inverse square root of the
determinant of the Laplacian, we find that it contributes to the
effective Euclidean worldsheet action
\be \label{SgaugeN} -\frac{1}{48\pi}\int
d^2\sigma\,\hat{e}\left(-\hat{h}^{mn}\p_m\phi\p_n\phi
+12k_+^2\hat{h}^{mn}\p_mX^+\p_nX^+\right). \ee
We can rewrite this contribution as a path integral over a bosonic
ghost-antighost system,
\be \frac{1}{\det
J_{\psi_-^+\epsilon}}=\int\CD\beta\,\CD\gamma\,\exp \left\{-\int
d^2\sigma\beta\frac{1}{F}D_-(F\gamma)\right\}. \ee
The antighost $\beta$ has the same conformal dimension as
$\psi_-^+$, or $(0,1/2)$, while the ghost field $\gamma$ has the
dimension of $\epsilon$, namely $(0,-1/2)$.  The path-integral
measure of the ghost and antighost fields, $\CD\beta\,\CD\gamma$, is
defined in the standard way, independently of $X^+$.  More
precisely, the standard measure on the fluctuations $\delta f$ of a
spin $j$ field $f$ is induced from the covariant norm
\be \label{snorm} \|\delta f\|^2=\int d^2\sigma e^{(2-2j)\phi}\delta
f^\ast\delta f, \ee
written here in conformal gauge $h_{mn}=e^{2\phi}\delta_{mn}$.

The ghost fields $\beta$ and $\gamma$ have a kinetic term suggesting that they
should be purely left-moving fields, but their conformal dimensions do not
conform to this observation.  It is natural to introduce the rescaled fields
\be
\label{rescgh}
\tilde\beta=\frac{\beta}{F},\qquad\tilde\gamma=F\gamma.
\ee
Because the conformal weight of $F$ is $(1/2,1/2)$, both $\tilde\beta$ and
$\tilde\gamma$ are now fields of conformal dimension $(1/2,0)$.  Moreover, in
terms of these rescaled fields, the classical ghost action takes the canonical
form of a purely left-moving first-order system of central charge $c=-1$,
\be
S_{\tilde\beta\tilde\gamma}=\int d^2\sigma\,\tilde\beta D_-\tilde\gamma.
\ee
However, the rescaling (\ref{rescgh}) has an
effect on the measure in the path integral.  In terms of the rescaled
variables, the originally $X^+$ independent measure acquires a non-canonical,
explicit $X^+$ dependence.  The new measure on $\tilde\beta$ and
$\tilde\gamma$ is induced from
\be
\label{xnorm}
\|\delta\tilde\beta\|^2=\int d^2\sigma e^\phi F^2\delta\tilde\beta^\ast
\delta\tilde\beta,\qquad
\|\delta\tilde\gamma\|^2=\int d^2\sigma e^\phi F^{-2}\delta\tilde\gamma^\ast
\delta\tilde\gamma.
\ee
In order to distinguish it from the standard $X^+$ independent
measure, we denote the measure induced from (\ref{xnorm}) by
$\CDX\tilde\beta\,\CDX\tilde\gamma$.  It is convenient to
replace
this $X^+$ dependent measure by hand with the standard measure
$\CD\tilde\beta \,\CD\tilde\gamma$ for left-moving fields
$\tilde\beta$, $\tilde\gamma$ of spin 1/2, defined with the use of
the standard norm (\ref{snorm}) that is independent of $X^+$.  In
order to do so, we must correct for the error by including the
corresponding Jacobian, which is precisely the $X^+$ dependent part
of $J_{\psi_-^+\epsilon}$.  Thus, we can now write
\bea \label{Sgauge} &&\frac{1}{\det
J_{\psi_-^+\epsilon}}=\int\CDX\tilde\beta\,\CDX\tilde\gamma
\,
\exp\left\{-\int d^2\sigma\,\tilde\beta D_-\tilde\gamma\right\}\\
&&\quad{}=\exp\left\{\frac{1}{48\pi}\int d^2\sigma\,\hat
e\left(12k_+^2\hat h^{mn}
\p_mX^+\p_nX^+\right)\right\}\int\CD\tilde\beta\,\CD\tilde\gamma\,
\exp\left\{-\int d^2\sigma\,\tilde\beta
D_-\tilde\gamma\right\}.\nonumber \eea
The canonical $\tilde\beta$, $\tilde\gamma$ ghosts correctly
reproduce the Liouville dependence of $J_{\psi_-^+\epsilon}$. Of
course, we now have a contribution to the effective action.  Written
in Minkowski signature, this correction becomes
\be \label{SgaugeMink} \frac{1}{4\pi}\int d^2\sigma\,\hat e\left(k_+^2
\hat h^{mn} \p_mX^+\p_nX^+\right). \ee

\subsection{$\tilde J$}

The evaluation of $J_{\psi^+_-\epsilon}$ is not the only calculation
that can contribute to the shift of the linear dilaton.  We also
need to analyze the determinant involved in the change of variables
that turns the gravitino and its conjugate field into manifestly
left-moving fields.  As we shall now show, this transformation also
contributes to the linear dilaton shift.

Consider the kinetic term between the conjugate pair of $\chi_{++}$
and $\psi_-^-$.  The relevant part of the path integral, written
here still in the Minkowski worldsheet signature, is
\be
\int\CD\chi_{++}\CD\psi_-^-\,\exp\left\{\frac{i\kappa V_-}{\pi}\int
d^2\sigma^\pm\left(\chi_{++}\p_-\psi_-^--k_+\chi_{++}\p_-X^+\psi_-^-\right)
\right\}.
\ee
This path integral would give the determinant of the operator $D_F$
acting on fields of spin $j=3/2$.  Wick rotating to Euclidean
signature and using (\ref{trivialdet}), we obtain for this
determinant
\be \label{gravdet} \exp\left\{-\frac{1}{48\pi}\int
d^2\sigma\,e\left(11\hat{h}^{mn}\p_m\phi\p_n\phi
+24k_+\hat{h}^{mn}\p_m\phi\p_nX^++
12k_+^2\hat{h}^{mn}\p_mX^+\p_nX^+\right)\right\}. \ee
As in the calculation of the Faddeev-Popov determinant above, it is
again useful to rescale the fields by an $X^+$ dependent factor. This
rescaling was in fact introduced  in Eqn.~(\ref{rescale}), repeated here for
convenience:
\be
\tilde\chi_{++}=F\chi_{++},\qquad\tilde\psi_-^-=\frac{\psi_-^-}{F}.
\ee
The first term in the conformal anomaly can be interpreted as due to
the purely left-moving conjugate pair of fields $\tilde\chi_{++}$
and $\tilde\psi_-^-$ with the standard $X^+$ independent measure
induced from (\ref{snorm}).  In order to reproduce correctly the
full determinant (\ref{gravdet}), we again have to compensate for
the $X^+$ dependence of the measure by including the rest of the
gravitino determinant (\ref{gravdet}) as an explicit conversion
factor.  Thus, the consistent transformation of fields includes the
measure change
\be \CDX\tilde\chi_{++}\,\CDX\tilde\psi_-^-= \tilde
J\,\CD\tilde\chi_{++}\,\CD\tilde\psi_-^-, \ee
with
\be \tilde J=\exp\left\{-\frac{1}{4\pi}\int
d^2\sigma\,e\left(2k_+\hat{h}^{mn}\p_m\phi\p_nX^++
k_+^2\hat{h}^{mn}\p_mX^+\p_nX^+\right)\right\}. \ee

Together with the contribution from the Faddeev-Popov determinant
(\ref{Sgauge}), we see that the contributions to the $\p_+X^+\p_-X^+$ term
in fact cancel, and we are left with the following one-loop correction to the
Euclidean action due to the measure factors,
\be \label{effact} \Delta S_E=\frac{1}{2\pi}\int
d^2\sigma\,e\,k_+h^{mn}\p_m\phi\p_nX^+, \ee
together with the bosonic superghosts of spin $1/2$ (and with the canonical
path integral measure) and central charge $c=-1$, plus the gravitino sector
consisting of the canonical conjugate pair $\tilde\chi_{++}, \tilde\psi_-^-$
also with the canonical measure and central charge $c=-11$.

Integrating (\ref{effact}) by parts and using the following
expression for the worldsheet scalar curvature in conformal gauge,
\be
eR=-2\delta^{mn}\p_m\p_n\phi,
\ee
we end up with
\be \label{dshiftE}
\Delta S_E=\frac{1}{4\pi}\int d^2\sigma\,e\,k_+X^+R.%
\ee
Rotating back to Minkowski signature, we find
\be
\label{dshift}
\Delta S=-\frac{1}{4\pi} \int d^2\sigma\,e\,k_+X^+R.
\ee
This term represents an effective shift in the dilaton by $V_+=k_+$.
The total central charge of the combined matter and ghost system,
with the one-loop correction (\ref{dshift}) to the linear dilaton
included, is zero.

\bibliographystyle{JHEP}
\bibliography{e8}

\providecommand{\href}[2]{#2}\begingroup\raggedright\begin{thebibliography}{10}

\bibitem{sen}
A.~Sen, {\it Tachyon dynamics in open string theory},  {\em Int. J. Mod. Phys.}
  {\bf A20} (2005) 5513--5656,
  [\href{http://xxx.lanl.gov/abs/hep-th/0410103}{{\tt hep-th/0410103}}].

\bibitem{headrick}
M.~Headrick, S.~Minwalla, and T.~Takayanagi, {\it Closed string tachyon
  condensation: An overview},  {\em Class. Quant. Grav.} {\bf 21} (2004)
  S1539--S1565, [\href{http://xxx.lanl.gov/abs/hep-th/0405064}{{\tt
  hep-th/0405064}}].

\bibitem{witten}
E.~Witten, {\it Instability of the Kaluza-Klein vacuum},  {\em Nucl. Phys.}
  {\bf B195} (1982) 481.

\bibitem{fh}
M.~Fabinger and P.~Ho\v{r}ava, {\it Casimir effect between world-branes in
  heterotic M-theory},  {\em Nucl. Phys.} {\bf B580} (2000) 243--263,
  [\href{http://xxx.lanl.gov/abs/hep-th/0002073}{{\tt hep-th/0002073}}].

\bibitem{hs1}
S.~Hellerman and I.~Swanson, {\it Cosmological solutions of supercritical
  string theory},  \href{http://xxx.lanl.gov/abs/hep-th/0611317}{{\tt
  hep-th/0611317}}.

\bibitem{hs2}
S.~Hellerman and I.~Swanson, {\it Dimension-changing exact solutions of string
  theory},  \href{http://xxx.lanl.gov/abs/hep-th/0612051}{{\tt
  hep-th/0612051}}.

\bibitem{hs3}
S.~Hellerman and I.~Swanson, {\it Cosmological unification of string theories},
   \href{http://xxx.lanl.gov/abs/hep-th/0612116}{{\tt hep-th/0612116}}.

\bibitem{hs4}
S.~Hellerman and I.~Swanson, {\it Charting the landscape of supercritical
  string theory},  \href{http://xxx.lanl.gov/abs/arXiv:0705.0980 [hep-th]}{{\tt
  arXiv:0705.0980 [hep-th]}}.

\bibitem{Karczmarek}
J.~L. Karczmarek and A.~Strominger, {\it Closed string tachyon condensation at
  c = 1},  {\em JHEP} {\bf 05} (2004) 062,
  [\href{http://xxx.lanl.gov/abs/hep-th/0403169}{{\tt hep-th/0403169}}].

\bibitem{hk1}
P.~Ho\v{r}ava and C.~A. Keeler, {\it Noncritical M-theory in 2+1 dimensions as
  a nonrelativistic Fermi liquid},  {\em JHEP} {\bf 07} (2007) 059,
  [\href{http://xxx.lanl.gov/abs/hep-th/0508024}{{\tt hep-th/0508024}}].

\bibitem{pixie}
P.~Ho\v{r}ava and C.~A. Keeler, {\it Closed-string tachyon condensation and the
  worldsheet super-Higgs effect},
  \href{http://xxx.lanl.gov/abs/arXiv:0709.2162 [hep-th]}{{\tt arXiv:0709.2162
  [hep-th]}}.

\bibitem{hw1}
P.~Ho\v{r}ava and E.~Witten, {\it Heterotic and Type I string dynamics from
  eleven dimensions},  {\em Nucl. Phys.} {\bf B460} (1996) 506--524,
  [\href{http://xxx.lanl.gov/abs/hep-th/9510209}{{\tt hep-th/9510209}}].

\bibitem{hw2}
P.~Ho\v{r}ava and E.~Witten, {\it Eleven-dimensional supergravity on a manifold
  with boundary},  {\em Nucl. Phys.} {\bf B475} (1996) 94--114,
  [\href{http://xxx.lanl.gov/abs/hep-th/9603142}{{\tt hep-th/9603142}}].

\bibitem{klt}
H.~Kawai, D.~C. Lewellen, and S.~H.~H. Tye, {\it Classification of closed
  fermionic string models},  {\em Phys. Rev.} {\bf D34} (1986) 3794.

\bibitem{shanta}
S.~P. De~Alwis and A.~T. Flournoy, {\it Closed string tachyons and
  semi-classical instabilities},  {\em Phys. Rev.} {\bf D66} (2002) 026005,
  [\href{http://xxx.lanl.gov/abs/hep-th/0201185}{{\tt hep-th/0201185}}].

\bibitem{dh}
L.~J. Dixon and J.~A. Harvey, {\it String theories in ten-dimensions without
  space-time supersymmetry},  {\em Nucl. Phys.} {\bf B274} (1986) 93--105.

\bibitem{forgacs}
P.~Forg\'{a}cs, Z.~Horv\'{a}th, L.~Palla, and P.~Vecserny\'{e}s, {\it Higher
  level Kac-Moody representations and rank reduction in string models},  {\em
  Nucl. Phys.} {\bf B308} (1988) 477.

\bibitem{lewellen}
D.~C. Lewellen, {\it Embedding higher level Kac-Moody algebras in heterotic
  string models},  {\em Nucl. Phys.} {\bf B337} (1990) 61.

\bibitem{elitzur}
S.~Elitzur and A.~Giveon, {\it Connection between spectra of nonsupersymmetric
  heterotic string models},  {\em Phys. Lett.} {\bf B189} (1987) 52.

\bibitem{difrancesco}
P.~Di~Francesco, P.~Mathieu, and D.~S\'en\'echal, {\it Conformal field theory},
   {\em {\rm Springer}} (1997).

\bibitem{chl}
S.~Chaudhuri, G.~Hockney, and J.~D. Lykken, {\it Maximally supersymmetric
  string theories in d < 10},  {\em Phys. Rev. Lett.} {\bf 75} (1995)
  2264--2267, [\href{http://xxx.lanl.gov/abs/hep-th/9505054}{{\tt
  hep-th/9505054}}].

\bibitem{cp}
S.~Chaudhuri and J.~Polchinski, {\it Moduli space of CHL strings},  {\em Phys.
  Rev.} {\bf D52} (1995) 7168--7173,
  [\href{http://xxx.lanl.gov/abs/hep-th/9506048}{{\tt hep-th/9506048}}].

\bibitem{bv}
N.~Berkovits and C.~Vafa, {\it On the uniqueness of string theory},  {\em Mod.
  Phys. Lett.} {\bf A9} (1994) 653--664,
  [\href{http://xxx.lanl.gov/abs/hep-th/9310170}{{\tt hep-th/9310170}}].

\bibitem{bastianelli}
F.~Bastianelli, {\it A locally supersymmetric action for the bosonic string},
  {\em Phys. Lett.} {\bf B322} (1994) 340--343,
  [\href{http://xxx.lanl.gov/abs/hep-th/9311157}{{\tt hep-th/9311157}}].

\bibitem{kunitomo}
H.~Kunitomo, {\it On the nonlinear realization of the superconformal symmetry},
   {\em Phys. Lett.} {\bf B343} (1995) 144--146,
  [\href{http://xxx.lanl.gov/abs/hep-th/9407052}{{\tt hep-th/9407052}}].

\bibitem{mca1}
I.~N. McArthur, {\it The Berkovits-Vafa construction and nonlinear
  realizations},  {\em Phys. Lett.} {\bf B342} (1995) 94--98,
  [\href{http://xxx.lanl.gov/abs/hep-th/9411150}{{\tt hep-th/9411150}}].

\bibitem{mca2}
I.~N. McArthur, {\it Gauging of nonlinearly realized symmetries},  {\em Nucl.
  Phys.} {\bf B452} (1995) 456--468,
  [\href{http://xxx.lanl.gov/abs/hep-th/9504160}{{\tt hep-th/9504160}}].

\bibitem{mcgreevy}
J.~McGreevy and E.~Silverstein, {\it The tachyon at the end of the universe},
  {\em JHEP} {\bf 08} (2005) 090,
  [\href{http://xxx.lanl.gov/abs/hep-th/0506130}{{\tt hep-th/0506130}}].

\bibitem{horowitz}
G.~T. Horowitz and E.~Silverstein, {\it The inside story: Quasilocal tachyons
  and black holes},  {\em Phys. Rev.} {\bf D73} (2006) 064016,
  [\href{http://xxx.lanl.gov/abs/hep-th/0601032}{{\tt hep-th/0601032}}].

\bibitem{aharony}
O.~Aharony and E.~Silverstein, {\it Supercritical stability, transitions and
  (pseudo)tachyons},  {\em Phys. Rev.} {\bf D75} (2007) 046003,
  [\href{http://xxx.lanl.gov/abs/hep-th/0612031}{{\tt hep-th/0612031}}].

\bibitem{flohr}
M.~Flohr, {\it Bits and pieces in logarithmic conformal field theory},  {\em
  Int. J. Mod. Phys.} {\bf A18} (2003) 4497--4592,
  [\href{http://xxx.lanl.gov/abs/hep-th/0111228}{{\tt hep-th/0111228}}].

\bibitem{gaberdiel}
M.~R. Gaberdiel, {\it An algebraic approach to logarithmic conformal field
  theory},  {\em Int. J. Mod. Phys.} {\bf A18} (2003) 4593--4638,
  [\href{http://xxx.lanl.gov/abs/hep-th/0111260}{{\tt hep-th/0111260}}].

\bibitem{orange}
P.~C. West, {\it Introduction to supersymmetry and supergravity},  {\em {\rm
  World Scientific, Singapore}} (1990) 2nd ed.

\bibitem{dhoker}
E.~D'Hoker and D.~H. Phong, {\it The geometry of string perturbation theory},
  {\em Rev. Mod. Phys.} {\bf 60} (1988) 917.

\bibitem{Kallosh}
R.~Kallosh and A.~Y. Morozov, {\it Green-Schwarz action and loop calculations
  for superstring},  {\em Int. J. Mod. Phys.} {\bf A3} (1988) 1943.

\bibitem{gerasimov}
A.~Gerasimov, A.~Morozov, M.~Olshanetsky, A.~Marshakov, and S.~L. Shatashvili,
  {\it Wess-Zumino-Witten model as a theory of free fields},  {\em Int. J. Mod.
  Phys.} {\bf A5} (1990) 2495--2589.

\end{thebibliography}\endgroup
\end{document}